\documentclass[11 pt]{article} \textheight=22 true cm
\usepackage{subfigure}
\usepackage{graphicx}
\usepackage{amsmath}
\usepackage{amssymb}
\usepackage{color}
\usepackage[colorlinks]{hyperref}
\usepackage{caption}
\usepackage{subcaption}
\usepackage{graphicx} 
\usepackage{enumerate}
\usepackage[utf8]{inputenc}
\usepackage{amsmath,amssymb}
\usepackage{caption}
\usepackage{subcaption}  
\usepackage{booktabs}
\usepackage{multirow}

\title{Emergent Universe Scenario in the Modified Chaplygin gas : Towards an Exact Solution and  Observational Constraints}
\author{D. Panigrahi\footnote{ Netaji Nagar Day College, 170/436 N.S. C. Bose Road, Regent Estate, Kolkata  700092, INDIA
\emph{and also} Relativity and Cosmology Research Centre, Jadavpur
University, Kolkata - 70032, India , e-mail:
dibyendupanigrahi@yahoo.co.in, dpanigrahi@nndc.ac.in }, S. Chatterjee \footnote{ New Alipore College (Retd.), Kolkata - 700053, India \emph{and
also} Relativity and Cosmology Research Centre, Jadavpur
University,
Kolkata - 700032, India, e-mail : chat\_sujit1@yahoo.com },  B. C. Paul\footnote{Department of Physics, University of North Bengal, Dist.-Darjeeling, PIN-734013, India, e-mail : bcpaul@associates.iucaa.in}}
\date{}
\begin{document}

\maketitle
\begin{abstract}

The Modified Chaplygin Gas (MCG) model is revisited to examine its capability to describe the complete cosmic evolution within a unified framework. The model is found to be consistent with observational constraints, including the CMB shift parameter, while also supporting standard structure formation alongside late-time cosmic acceleration. Due to the highly nonlinear nature of the governing field equations, exact analytical solutions in cosmic time are generally not attainable. To overcome this, we adopt a first-order approximation, which enables us to obtain an analytical expression for the scale factor and determine the corresponding flip time. The effective equation-of-state parameter evolves from a positive value in the early matter-dominated era to $w_{\mathrm{eff}} \approx -1$ at late times, in agreement with observations. In addition, a solution in the phantom regime exhibits an emergent, non-singular behavior, where the universe originates from a finite minimum scale factor and smoothly evolves toward a late-time de~Sitter--like phase, approaching the $\Lambda$CDM limit. The emergent solution is shown to be geodesically complete and free from curvature singularities, as confirmed by the regular behavior of the Kretschmann scalar. We also perform a CMB analysis for all the considered cases, and in each case the obtained values of $R_{\mathrm{th}}$ and $\chi^2_{\mathrm{CMB}}$ remain consistent with observations within the $1\sigma$ confidence level. Furthermore, the combined Hubble-$57$ and CMB analysis leads to significantly improved constraints on the model parameters. Moreover, the obtained solutions are independently verified using the Raychaudhuri equation, providing a consistency check that further reinforces the robustness of the methodology. Overall, the results demonstrate that the MCG model provides a self-consistent and observationally viable description of cosmic evolution without requiring modifications to general relativity.

\end{abstract}

KEY WORDS : cosmology;  accelerating universe; chaplygin gas \\

 PACS :   04.20,  04.50 +h
\bigskip
\section{ Introduction}

Following the discovery of high-redshift Type Ia supernovae~\cite{res}, it became evident that within the framework of the standard Friedmann–Robertson–Walker (FRW) cosmology —assuming homogeneity and isotropy —the universe is presently undergoing an accelerated phase of expansion. Observations indicate that baryonic matter contributes only about four percent of the total energy density, while the remaining components are of non-baryonic type. Independent evidence from Cosmic Microwave Background Radiation (CMBR) measurements~\cite{spe} also supports this accelerated expansion, thereby motivating extensive theoretical and observational efforts to understand its origin.

The fundamental challenge lies in identifying the mechanism responsible for this late-time acceleration. Broadly, two main theoretical approaches have been proposed to address the late acceleration: (i) modifications of Einstein’s general theory of relativity, and (ii) the introduction of a new cosmic fluid—often identifying an evolving cosmological constant or a dynamical scalar field such as quintessence. However, both the approaches face significant theoretical and conceptual issues. The phenomenon is generally attributed to an exotic component of the cosmic fluid, known as \textit{dark energy} (DE), which accounts for nearly three-fourths of the total energy density of the universe. Another crucial yet equally elusive component is \textit{dark matter} (DM), which provides the necessary gravitational pull to explain galaxy formation and large-scale structure.

Despite numerous attempts, identifying  the true nature of DE for the first case remains one of the most profound unresolved problem in cosmology. Over the past few decades, a wide range of DE models have been explored, including quintessence~\cite{sam}, phantom fields~\cite{cal}, holographic dark energy~\cite{li}, string-inspired models~\cite{lm}, Born–Infeld condensates~\cite{eli}, modified gravity theories~\cite{sha}, and inhomogeneous cosmological scenarios~\cite{kra}. Comprehensive reviews of these models are available in the literature~\cite{bamb}.

Although the evolving cosmological constant ($\Lambda$) provides a simplest explanation for the observed acceleration in the second case, it suffers from two major theoretical challenges: (i) the \textit{fine-tuning problem}, since the observed value of $\Lambda$ is vastly smaller than that predicted by quantum field theory, and (ii) the \textit{cosmic coincidence problem}, arising from the near equality of the present-day matter and vacuum energy densities despite their different evolutionary histories. While the $\Lambda$CDM model remains consistent with observations, these issues diminish its theoretical appeal. Attempts to generalize $\Lambda$ to a variable form introduces further complications~\cite{cop, dp06}, while alternative frameworks such as brane-world cosmology~\cite{ish, dp11, dp09} is also considered to invoke extra spatial dimensions that remain speculative.

However, among the alternative approaches, the \textit{Chaplygin Gas} (CG) model has emerged as an intriguing candidate. It is characterized by the equation of state (EoS)
\begin{equation}
p = -\frac{B}{\rho},
\label{eq:1a}
\end{equation}
where $\rho$ and $p$ denote the energy density and pressure, respectively, and $B$ is a positive constant~\cite{kam}. Although the CG model successfully explains supernova data, it fails to reproduce the observed structure formation and exhibits unphysical oscillations in the matter power spectrum~\cite{san}.
Thus  CG is genaralized,  assuming a  \textit{Generalized Chaplygin Gas} (GCG) which is
\begin{equation}
p = -\frac{B}{\rho^{\alpha}},
\label{eq:1b}
\end{equation}
where $0 < \alpha < 1$~\cite{bent}. The GCG model can effectively reproduce the background dynamics of a homogeneous and isotropic universe and mimic the $\Lambda$CDM model at late times. However, it suffers from strong oscillations or instabilities in the matter power spectrum unless it effectively reduces to $\Lambda$CDM~\cite{ame}.
A further refinement in the EoS named,  the \textit{Modified Chaplygin Gas} (MCG) model~\cite{bena, deb1,dp15} was proposed. MCG introduces an additional term linear in $\rho$, leading to an EoS of two different parameters ($\alpha$ and $A$),
\begin{equation}
p = A\rho - \frac{B}{\rho^{\alpha}}.
\label{eq:1c}
\end{equation}
The MCG model  provides a unified description of dark matter and dark energy within a single fluid framework. It exhibits a radiation-dominated behavior at high densities (for $A = \tfrac{1}{3}$), transitions through a matter-dominated phase, and asymptotically approaches a cosmological constant–like behavior at low densities. Thus, a universe filled with MCG  consistently evolves from  radiation epoch to the present dark energy–dominated phase, finally entering in to  the $\Lambda$CDM limiting case.
A number of studies in the literature reveled various aspects of the MCG model,  Wu \textit{et al.}~\cite{yu} explored its dynamical properties, Bedran \textit{et al.}~\cite{bed} examined the temperature evolution in its presence, and later works demonstrated its consistency with perturbative analyses~\cite{costa,ya} and spherical collapse scenarios~\cite{fab} where Febris et al studied and ruled out the constant $A$.
In the present work, we revisit a spatially flat Friedmann–Robertson–Walker (FRW) universe filled with a Modified Chaplygin Gas (MCG) and perform a detailed cosmological analysis to constrain its model parameters making use of observations. Using the standard $\chi^2$-minimization technique, we determine the best-fit values of the MCG parameters from the Hubble–57 dataset and employ these results to investigate the dynamical evolution of the universe.

It is important to note that the key Eq.~\eqref{eq:8} does not admit an explicit analytic solution for the scale factor with a known functional form.  Consequently, the time evolution of several cosmological quantities—such as evolution of the scale factor, the flip time etc. — cannot be obtained in closed form. To overcome this limitation in order to explore the temporal behavior of these quantities, we adopt an alternative analytical approach that enables us to study the underlying dynamics more transparently.
Furthermore, the concept of an emergent universe—a cosmological scenario where the universe is eternally existing and originates from a quasi-static state before entering an inflationary expansion—has been widely investigated in recent years~\cite{eli1, eli2, sm1}. Such model, consistent with the classical Lemaître–Eddington framework, offer an elegant resolution to the initial singularity problem inherent in the standard Big Bang cosmology.
A physically consistent emergent universe model  naturally addresses several conceptual issues of the Big Bang scenario. For instance,  Mukherjee \textit{et al.}~\cite{sm} obtained an emergent-type solution by setting a negative $\alpha$ ($\alpha = -\tfrac{1}{2}$) in Eq.~\eqref{eq:1c}, whereas Dutta \textit{et al.}~\cite{sd} claimed that such non-singular model is not possible  within the conventional MCG framework. In the paper, we demonstrated that the analysis based on Eq.~(57) reveals a new class of emergent-type cosmology arising from the MCG equation of state under the first-order approximation, which satisfies the conditions $0<A<\frac{1}{3}$ and $0<\alpha<1$. Emergent Universe (EU) model is an acceptable cosmological model of the universe which is singularity free. Stability and low CMB analysis of the EU in general relativity \cite{5a,5b} and in Jordan frame \cite{6a,6b,6c,6d} are analyzed and found to work well.

The paper is organized as follows. In Sec.~2, we present the fundamental field equations derived from Einstein’s equations. Sec.~3 is devoted to the mathematical formulation of the model, leading to a hypergeometric-type solution. In this section, we also derive expressions for the deceleration parameter, the effective equation of state, and the redshift corresponding to the transition (flip) time. It is shown that the model asymptotically approaches the $\Lambda$CDM behavior at late cosmic times, and its dynamical evolution is illustrated through graphical analyses. We also discuss the possibility of structure formation within the MCG framework by examining the existence of a prolonged quasi-matter-dominated phase that is favorable for the growth of density perturbations. In Sec.~4, we highlight an important aspect of our analysis by considering a first-order approximation of the key dynamical equation. This approximation allows us to obtain an exact analytical expression for the scale factor, and a detailed investigation of the flip time is carried out both analytically and graphically. In Sec.~5, we demonstrate that the model admits an emergent universe scenario in the phantom regime for the integration constant $c < 0$. The evolution of the scale factor, effective equation of state, time-varying deceleration parameter, and energy density is analyzed in detail. We also examine the geodesic completeness and the absence of curvature singularities through the behavior of the Kretschmann scalar. Furthermore, the model parameters are constrained using the Hubble-$57$ dataset via a standard $\chi^2$ minimization technique. The CMB shift parameter is also calculated, followed by a joint Hubble-$57$ and CMB analysis. In Sec.~6, we discuss the physical implications of the results within the framework of the Raychaudhuri equation. Finally, Sec.~7 summarizes the main findings and presents the concluding remarks.

\section{ Field Equations}
We consider a spherically symmetric homogeneous spacetime given by
\begin{equation}\label{eq:1}
  ds^{2}= dt^{2}- a^2(t)~\{dr^{2}+r^{2}\left( d\theta^{2}+\sin^{2}\theta d\phi^{2} \right) \},
\end{equation}
where the scale factor, $a(t)$  depends on time only.

The
energy momentum tensor is given by

\begin{equation}\label{eq:2}
T^{\mu}_{\nu} = (\rho + p)\delta_{0}^{\mu}\delta_{\nu}^{0} - p\delta_{\nu}^{\mu},
\end{equation}
where $\rho(t)$ is the matter density  and $p(t)$ the isotropic
 pressure.
 \vspace{0.1 cm}
 Here we consider a comoving coordinate system such that $ u^{0}=1, u^{i}= 0 ~(i = 1, 2,3)$ and $g^{\mu
\nu}u_{\mu}u_{\nu}= 1$ where $u_{i}$ is the 4- velocity.

The independent field equations for the metric  Eq.~\eqref{eq:1} and the energy
momentum tensor Eq.~\eqref{eq:2} are given by

\begin{equation} \label{eq:3}
3\frac{\dot{a}^2}{a^2} = 3H^2 =\rho,
\end{equation}
\begin{equation}\label{eq:4}
 2 \frac{\ddot{a}}{a} + \frac{\dot{a}^2}{a^2} = 2(\dot{H} + H) = - p,
\end{equation}
where $H$ is the Hubble parameter. Now from the conservation law for homogeneous model

\begin{equation} \label{eq:5}
\dot{\rho} + 3 H (\rho + p) = 0.
\end{equation}

At this stage, we assume that the cosmic fluid is governed by the Modified Chaplygin Gas (MCG) equation of state, as given in Eq.~\eqref{eq:1c}. Using Eqs~\eqref{eq:1c} \& \eqref{eq:5} and performing straightforward algebraic manipulations, we obtain the following expression for the energy density

\begin{equation}\label{eq:7}
\rho = \left[ \frac{B}{(1+A)} +
c (1+z)^{3(1+\alpha)(1+A)}\right]^{\frac{1}{1+\alpha }},
\end{equation}
where $c$ is an integration constant. Plugging in the expression of $\rho$ from Eqs~ \eqref{eq:3} and \eqref{eq:7} we
finally get

\begin{equation}\label{eq:8}
3 \frac{\dot{a}^{2}}{a^{2}} =3H^2 = \left[ \frac{B}{(1+A)} + c (1+z)^{3(1+\alpha)(1+A)}\right]^{\frac{1}{1+\alpha }}.
\end{equation}

 \vspace{0.1 cm}
\section{Cosmological dynamics}
The dynamics of the MCG model in a 4D framework have been studied extensively~\cite{bena, deb1}, with perturbative analyses highlighting several generic features~\cite{costa}. Notably, Fabris \emph{et al.}~\cite{fab} used perturbative methods to compare the model with observational data, concluding from power spectrum analyses that $A < 10^{-6}$, effectively reducing the model to the GCG and almost ruling out the MCG. Furthermore, recent supernova data suggest that negative values of $\alpha$ are favored~\cite{deng}, though such values may imply imaginary sound speeds and potential instabilities~\cite{batista,fabris,col}. Conversely, it has been argued that $\alpha > 1$ is also possible~\cite{lu}, though such cases risk superluminal sound speeds.

\vspace{0.5 cm}

\subsection{\textbf{Square speed of sound:}}
The adiabatic sound speed squared in the Modified Chaplygin Gas (MCG) model, defined by the Eq.~\eqref{eq:1c} is given by
\begin{equation}\label{eq:8a}
c_s^2 = \frac{\partial p}{\partial \rho} = A + \alpha \frac{B}{\rho^{\alpha+1}}.
\end{equation}
For physical viability, the sound speed must remain real, positive, and subluminal, i.e.,   $ 0 \leq c_s^2 \leq 1$.
\begin{enumerate}[(i)]
\item Early Universe ($\rho \gg 1$):
In the high-density regime, the second term in the equation of state is negligible, so the effective sound speed approaches
$c_s^2 \to A$. Thus, the parameter $A$ governs the early-time behavior of the fluid. To mimic radiation- or dust-like matter, one typically requires $ 0 \leq A \leq \frac{1}{3}$, which ensures consistency with nucleosynthesis and structure formation.

\item  Late Universe ($\rho \ll 1$):
As the density decreases with cosmic expansion, the second term grows, and  $ c_s^2 \simeq \alpha \frac{B}{\rho^{\alpha+1}} $.
This can diverge and exceed unity unless $\alpha$ is sufficiently small. Rearranging the causality condition yields $
\alpha \leq \frac{(1 - A)\rho^{\alpha+1}}{B}$. Hence, the upper bound on $\alpha$ is density-dependent and becomes increasingly restrictive as $\rho \to 0$. This feature reflects the MCG’s transition from matter-like to dark-energy–dominated behavior, with the equation of state naturally softening at low densities.
\end{enumerate}

To ensure stability and consistency with cosmology, the following parameter ranges are commonly imposed:  $ 0 \leq A \leq \tfrac{1}{3}, \quad 0 \leq \alpha < 1, \quad B > 0$.
These conditions guarantee that the sound speed remains real, subluminal, and stable under perturbations throughout cosmic evolution. In our analysis, we have consistently imposed and maintained these allowed parameter ranges.

 An exact closed-form solution for the scale factor $a(t)$ from Eq.~\eqref{eq:8}  is difficult to obtain, as the integration results in an elliptic form. However, the equation yields valuable information in certain limiting cases as discussed below.
\subsection{\textbf{Deceleration Parameter:}}

 During the early phase of cosmological evolution, with a relatively small scale factor $a(t)$, the second term in Eq.~\eqref{eq:8} becomes dominant. This regime has already been addressed in detail in the literature \cite{bena,deb1}, so we only summarize the main point here. From the expression of the deceleration parameter $q$, it follows that

\begin{equation}\label{eq:9}
q = -\frac{1}{H^{2}}\frac{\ddot{a}}{a}= \frac{d}{dt} \left(H^{-1}
\right) - 1 = \frac{1}{2} + \frac{3}{2}\frac{p}{\rho} = \frac{1 +3A}{2}- \frac{3B}{2}\frac{1}{\rho^{\alpha +1}},
\end{equation}
which, by Eq.~\eqref{eq:7}, gives

\begin{equation}\label{eq:10}
q = \frac{1 +3A}{2}- \frac{3B}{2}\left[\frac{B}{1+A}+ C (1+z)^{3(1+\alpha)(1+A)}\right]^{-1}.
\end{equation}
To constrain the parameters, we consider $\Omega_m = \frac{c}{\rho_0^{1+\alpha}} $ and $\frac{B}{1+A} = \rho_0^{1+\alpha} (1- \Omega_m)$  and we can express the Eq.~\eqref{eq:10} as

\begin{equation}\label{eq:11}
q = \frac{1 +3A}{2}- \frac{3}{2} (1+A) \frac{1 - \Omega_m}{\Omega_m}  \left[ \frac{1 - \Omega_m}{\Omega_m}  + (1+z)^{3(1+\alpha)(1+A)}\right]^{-1}.
\end{equation}

At the flip time $q = 0$, the redshift parameter at the flip time $z_f$  ~\cite{per, res1} is given by from Eq.~\eqref{eq:10} as
\begin{equation}\label{eq:12}
z_f = \left[\frac{2B}{C} \frac{1} {(1+A)(1+3A)}\right]^{\frac{1}{3(1+\alpha)(1+A)}} - 1,
\end{equation}
and from Eq.~\eqref{eq:11}, we get
\begin{equation}\label{eq:13}
z_f = \left[\frac{1- \Omega_m}{\Omega_m} \frac{2}{1 + 3A} \right]^{\frac{1}{3(1+A)(1+\alpha)}} -1,
\end{equation}
where $z_f$ signifies the sign change of the deceleration parameter. For the universe to be accelerating at the present epoch (i.e., at $z = 0$), we require $z_f > 0$, which gives from Eq.~\eqref{eq:10}  as (i) $B > \frac{c}{2} (1+3A)(1+A)$ and from Eqn.~\eqref{eq:11} as (ii) $\Omega_m < \frac{2}{3(1+A)}$, since $0 \leq A \leq \frac{1}{3} $, this is consistent with the current observational constraint on $\Omega_m$.

\begin{figure}[ht]
\begin{center}
    \includegraphics[width=7.2 cm]{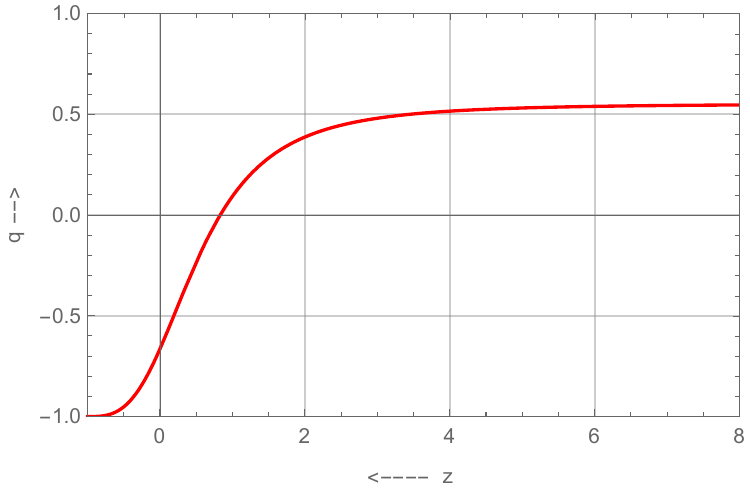}
      \caption{
  \small\emph{The variations of $q$ and $z$. }\label{qz}
    }
\end{center}
\end{figure}
As the universe expands, the energy density $\rho$ decreases with time, causing the second term in Eq.~\eqref{eq:9} to increase. This indicates the occurrence of a transition (flip) when the density reaches a critical value $\rho = \rho_{flip} = \left(\frac{3B}{1 + 3A}\right)^{\frac{1}{1 + \alpha}}$.
Since $A  > 0 $ and $\alpha > 0$ correspond to a lower density at the transition epoch, they favor a later onset of cosmic acceleration. This indicates that the universe remains in the decelerating phase for an extended period before the dark-energy–like behavior of the Modified Chaplygin Gas drives acceleration.

\vspace{0.5cm}

The behavior of the deceleration parameter $q$ at different epochs of cosmic evolution can be summarized as follows.
\begin{enumerate}[(i)]
    \item  At the early stage, when the redshift $z$ is very high  the  Eq.~\eqref{eq:11} reduces to

\begin{equation}\label{eq:14}
q = \frac{1 +3A}{2}.
\end{equation}
For the FRW model, during the radiation-dominated era, $q = 1$, which from Eq.~\eqref{eq:14} yields $A = \frac{1}{3}$.

 \item At present epoch ($z = 0$),  the Eq.~\eqref{eq:11} reduces to

\begin{equation}\label{eq:15}
q = \frac{1 +3A}{2} - \frac{3}{2}(1+A)(1-\Omega_m).
\end{equation}
Since $\Omega_m < \tfrac{2}{3(1+A)}$, we obtain $q < 0$, which corresponds to an accelerating universe.

\item At the late stage of cosmic evolution, Eq.~\eqref{eq:11} reduces to $q = -1$, representing $\Lambda$CDM.
\end{enumerate}
By setting $A = 0$, the present model reduces to the Generalized Chaplygin Gas (GCG) form.
In this limit, the resulting expressions are identical to those discussed in our earlier work~\cite{dp25}.

Fig-\ref{qz} illustrates the variation of the deceleration parameter $q$ with redshift $z$ in the framework of the Modified Chaplygin Gas (MCG) model. The universe transits from early deceleration ($q>0$) to late-time acceleration ($q<0$) at the flip redshift $z_f$, showing a smooth evolution without separate dark matter or dark energy components.

\vspace{0.2 cm}

\subsection{\textbf{Effective equation of state:}}

From Eqs.~\eqref{eq:1c} and \eqref{eq:7}, the effective equation of state parameter $w_{\text{eff}}$ is obtained as

\begin{equation}\label{eq:16}
w_{\text{eff}} = \frac{p}{\rho} = A - \frac{B}{\rho^{1+\alpha}} =  A - \frac{(1+A)(1-\Omega_m)}{1 - \Omega_m + \Omega_m(1+z)^{3(1+A)(1+ \alpha)}}.
\end{equation}
The effective equation of state parameter $w_{\text{eff}}$ exhibits distinct behavior across different cosmic epochs as follows:

\begin{enumerate}[(i)]

\item Early Universe: In the high-redshift limit, using Eq.~\eqref{eq:16}, we find $ w_{\text{eff}} = A$, corresponding to a early phase of the universe.

\item Dust dominated Epoch: At the dust-dominated epoch, the effective equation-of-state parameter becomes $w_{\text{eff}} = 0$, which yields
\begin{equation}\label{eq:16a}
z_{\text{dust}} = \left( \frac{1 - \Omega_m}{A \Omega_m} \right)^{\frac{1}{3(1+A)(1+\alpha)}} - 1.
\end{equation}
A comparison with the redshift at the flip time, $z_f$, shows that $z_{\text{dust}} > z_f$, implying that the transition to the accelerated expansion occurs after the dust-dominated era, which is consistent with the standard cosmological timeline of the evolution of the universe.

\item Present Epoch: In the present epoch ($z = 0$), Eq.~\eqref{eq:16} simplifies to
\begin{equation}\label{eq:17}
w_{\text{eff}} = -1 + \Omega_m(1+A).
\end{equation}
Since the maximum value of $\Omega_m$ is   $(\Omega_m)_{\text{max}} = \tfrac{2}{3(1+A)}$, this yields $(w_{\text{eff}} )_{max}= - \tfrac{1}{3}$. Therefore, at $z = 0$, we obtain $-1 < w_{\text{eff}} < - \frac{1}{3}$, which corresponds to an accelerating universe.

\item Late Universe: At the late stage of cosmic evolution, the Eq.~\eqref{eq:16} reduces to $w_{\text{eff}} = -1$,
indicating that the model asymptotically approaches a state similar to a cosmological constant, effectively representing the scenario $\Lambda$CDM.

\end{enumerate}

Thus, the model smoothly interpolates between the early phase and a cosmological constant--like late-time phase, providing a consistent description of cosmic evolution.

\begin{figure}[ht]
    \centering
    \subfigure[\small\emph{$ -1 \leq z \leq 12 $}]{
        \includegraphics[width=6 cm]{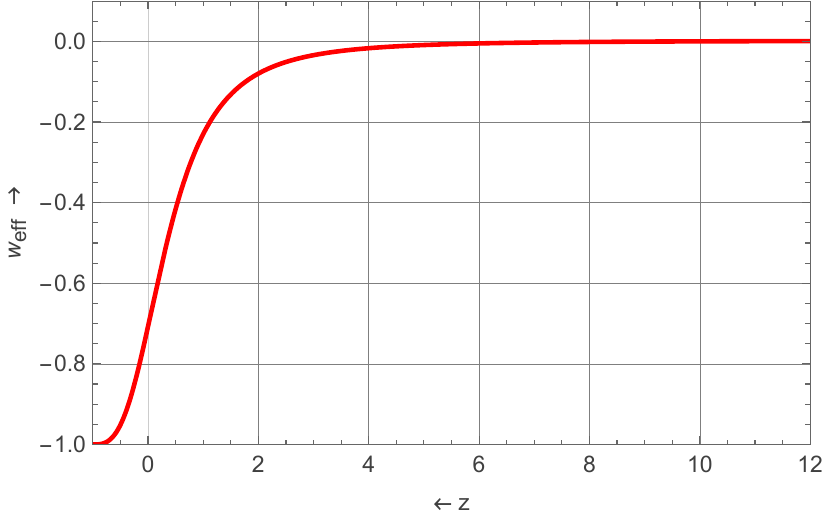}
        \label{wz1}
    }
    ~~~
    \subfigure[\small\emph{$3.5 \leq z \leq 12 $}]{
        \includegraphics[width=6 cm]{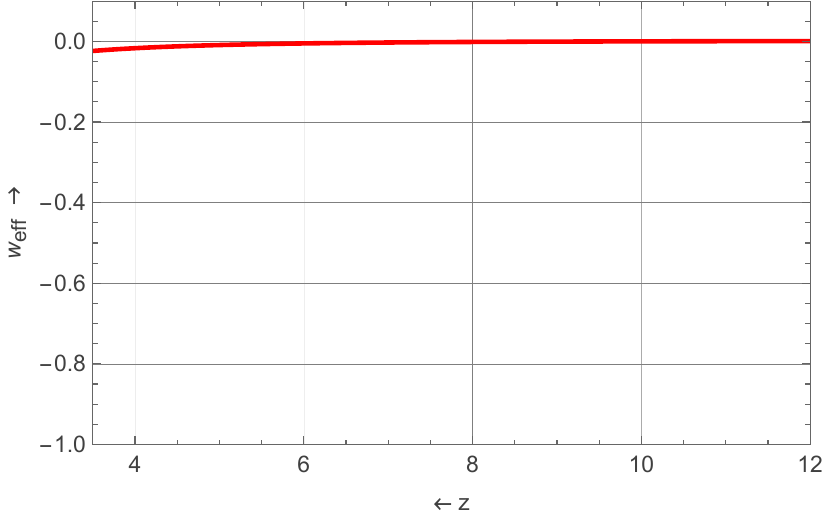}
        \label{wz2}
    }
    \caption[\small Optional caption for list of figures]{
        \small\emph{The variations of $w_{\text{eff}}$ and $z$}
    }
    \label{wz}
\end{figure}

It may be noted that on setting $A = 0$, the effective equation of state reduces to that of the Generalized Chaplygin Gas (GCG) model,
for which the derived expressions are consistent with those obtained in our earlier work~\cite{dp25}.

Fig.~\ref{wz1} illustrates the evolution of the effective equation-of-state parameter $w_{\mathrm{eff}}$ over the redshift range $-1 \leq z \leq 12 $ in the Modified Chaplygin Gas (MCG) model. At higher redshifts, $w_{\mathrm{eff}}>0$ corresponds to a decelerated, matter-like phase, whereas near the present epoch, $-1<w_{\mathrm{eff}}< - \frac{1}{3}$ indicates an accelerated expansion driven by an effective dark-energy component. At late times, $w_{\mathrm{eff}}\rightarrow -1$, asymptotically approaching the $\Lambda$CDM limit.

To highlight the behavior during the intermediate epoch, Fig.~\ref{wz2} presents an enlarged view of the redshift interval $3.5 \leq  z  \leq  12$. It is evident that $w_{\mathrm{eff}}$ remains very close to zero throughout this range, indicating an extended quasi-matter-dominated phase. Such a prolonged matter-like era is favorable for the growth of density perturbations and structure formation before the onset of the late-time accelerated expansion. Thus, the MCG model smoothly interpolates between the early matter-like phase and a cosmological constant--like late-time phase, providing a consistent description of cosmic evolution.

Next, we carry out an observational constraint analysis to determine the model parameters within the Modified Chaplygin Gas (MCG) framework.

\subsection{Observational Constraints}

In this subsection, we first determine the model parameters using the Hubble-$57$~\cite{dp25}  dataset. Subsequently, we perform a joint analysis by combining the Hubble-$57 $ dataset with the CMB shift parameter in order to obtain tighter constraints on the model parameters.

\subsubsection{Constraints from Hubble-$57$ dataset}
In this section, the Hubble-$57$ data~\cite{dp25} will be utilized to analyze the cosmological model by estimating the constraints imposed on the model parameters. Using the present value of the scale factor normalized to unity, \emph{i.e.}, $a = a_0 = 1$, we obtain a relation between the Hubble parameter and the redshift parameter $z$.
If $\rho_0$ denotes the density at the present epoch, then the well-known density parameter is written as $\Omega_m = \frac{c}{\rho_0^{1+\alpha}}$~\cite{seth}. Now, using Eq.~\eqref{eq:7}, we can express the three-dimensional spatial matter density as

\begin{equation}\label{eq:18}
\rho = \rho_0 \left \{1-\Omega_m + \Omega_m \left(1+z \right)^{3(1+A)(1+\alpha)} \right \}^{\frac{1}{1+\alpha}}.
\end{equation}

\begin{figure}[ht]
\begin{center}
  \includegraphics[width=10cm]{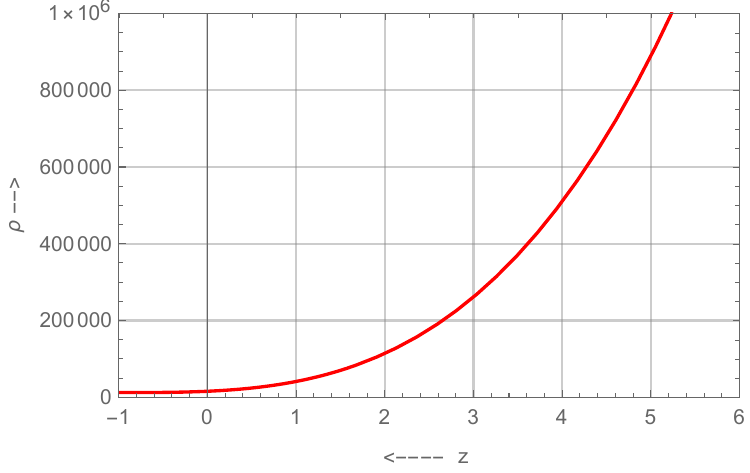}
  \caption{
  \small\emph{The variation of $\rho$ with $z$ is shown in this figure.    }\label{rz}
    }
\end{center}
\end{figure}
\begin{figure}[ht]
    \centering
    \subfigure[\small\emph{Best fit graph using Hubble $57$ data}]{
        \includegraphics[width=5.5cm]{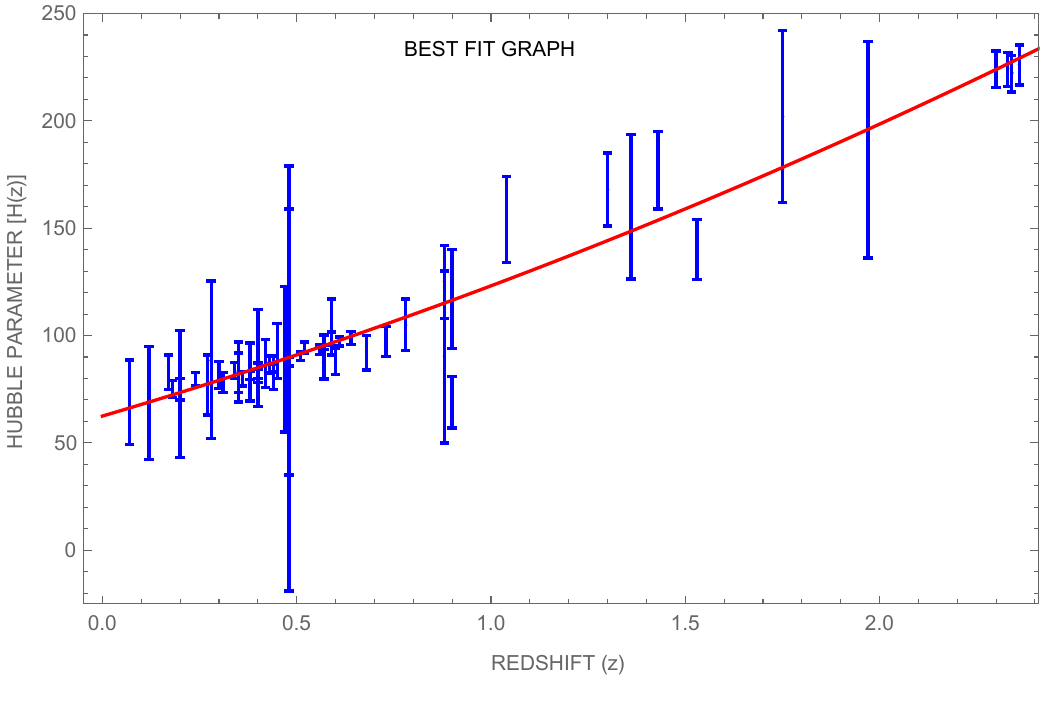}
        \label{hz1}
    }
    ~~~
    \subfigure[\small\emph{Best fit graph with Theoretical graph}]{
        \includegraphics[width=5.5cm]{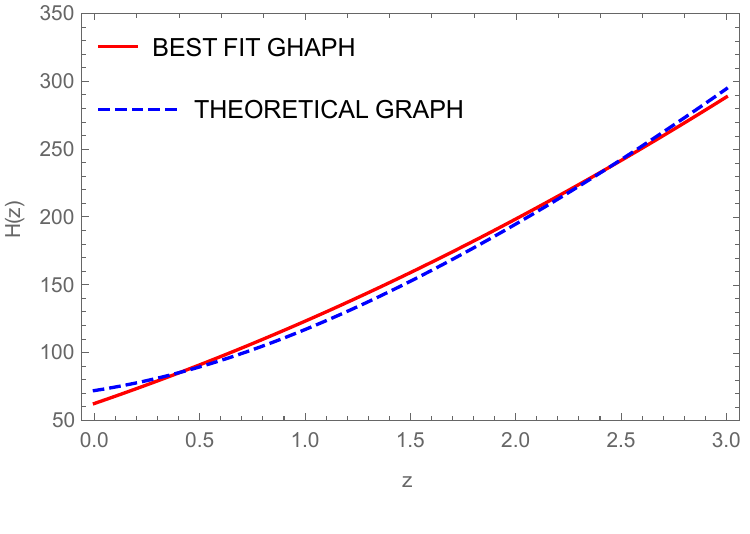}
        \label{hz2}
    }
    \caption[\small Optional caption for list of figures]{
        \small\emph{$H(z)$ vs $z$}
    }
    \label{fig:hz}
\end{figure}

The Fig.-\ref{rz} illustrates the evolution of the universe's density as a function of redshift, $z$. It shows that the matter density decreases as $z$ decreases, consistent with expectations from cosmological models where matter dilutes as the universe expands. Now the Hubble parameter
\begin{equation}\label{eq:19}
 H(z) = H_0 \left \{1-\Omega_m + \Omega_m \left(1+z \right)^{3(1+A)(1+\alpha)} \right \}^{\frac{1}{2(1+\alpha)}},
\end{equation}
where $H_0 = \left(\frac{\rho_0}{3} \right)^{\frac{1}{2}}$ represents the present value of the Hubble parameter. The Eq.~\eqref{eq:19} describes the evolution of the Hubble parameter $H(z)$ as a function of the redshift parameter $z$. In Fig.-\ref{hz1}, we present a best-fit curve of the redshift $z$ against the Hubble parameter $H(z)$ using the Hubble-$57$ data points. Furthermore, in Fig.-\ref{hz2}, we compare the best-fit graph with the graph obtained from Eq.~\eqref{eq:19}. These two graphs nearly coincide throughout the evolution, indicating that the behavior of our model is in good agreement with the observational data.

The apparently small uncertainty of the measurement naturally increases its weightage in
estimating $\chi^2$ statistics. We define here the   $\chi^2$ as
\begin{equation}\label{eq:20}
\chi_{H}^2 = \sum_{i=1}^{30} \frac{[H^{obs}(z_i) - H^{th} (z_i, H_0, \theta]^2}{\sigma^2_H(z_i)}.
\end{equation}
Here, $H^{\text{obs}}$ represents the observed Hubble parameter at $z_i$, and $H^{\text{th}}$ is the corresponding theoretical Hubble parameter given by Eq.~\eqref{eq:19}. Additionally, $\sigma_H(z_i)$ denotes the uncertainty for the $i$-th data point in the sample, and $\theta$ represents the model parameter. In this work, we utilize the observational $H(z)$ dataset, which consists of $57$ data points spanning the redshift range $0.07 \leq z \leq 2.36$~\cite{dp25}, extending beyond the redshift range covered by type Ia supernova observations. It is important to note that the confidence levels $1\sigma$ ($68.3\%$), $2\sigma$ ($95.4\%$), and $3\sigma$ ($99.7\%$) correspond to $\Delta \chi^2$ values of $2.3$, $6.17$, and $11.8$, respectively, where $ \Delta \chi^2 = \chi^2(\theta) - \chi^2(\theta^*) $ and $\chi^2_m$ is the minimum value of $\chi^2$. An important quantity used in the data fitting process is
\begin{equation}\label{eq:20a}
\overline{\chi^2} =  \frac{\chi_m^2}{dof},
\end{equation}
where, the subscript {\it dof} represents the degrees of freedom, defined as the difference between the total number of observational data points and the number of free parameters. If $\frac{\chi_m^2}{\textit{dof}} \leq 1$, it indicates a good fit, implying that the observed data are consistent with the considered model.

\begin{figure}[h!]
    \centering 
    \begin{tabular}{l l}
        \parbox{2in}{%
            \includegraphics[width=2in,height=2in]{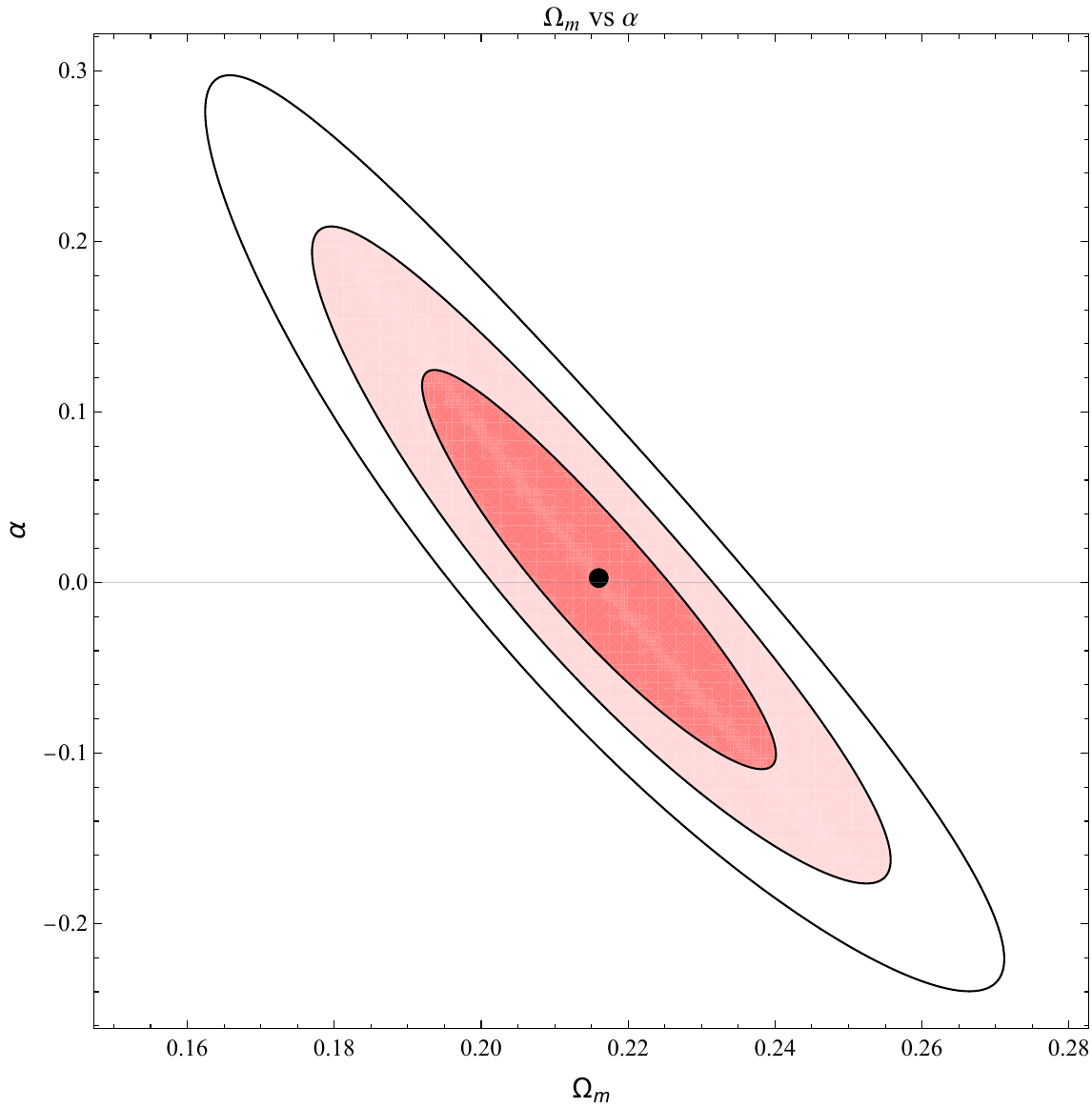}%

        } &
        \parbox{2in}{%
           \includegraphics[width=2in,height=1in]{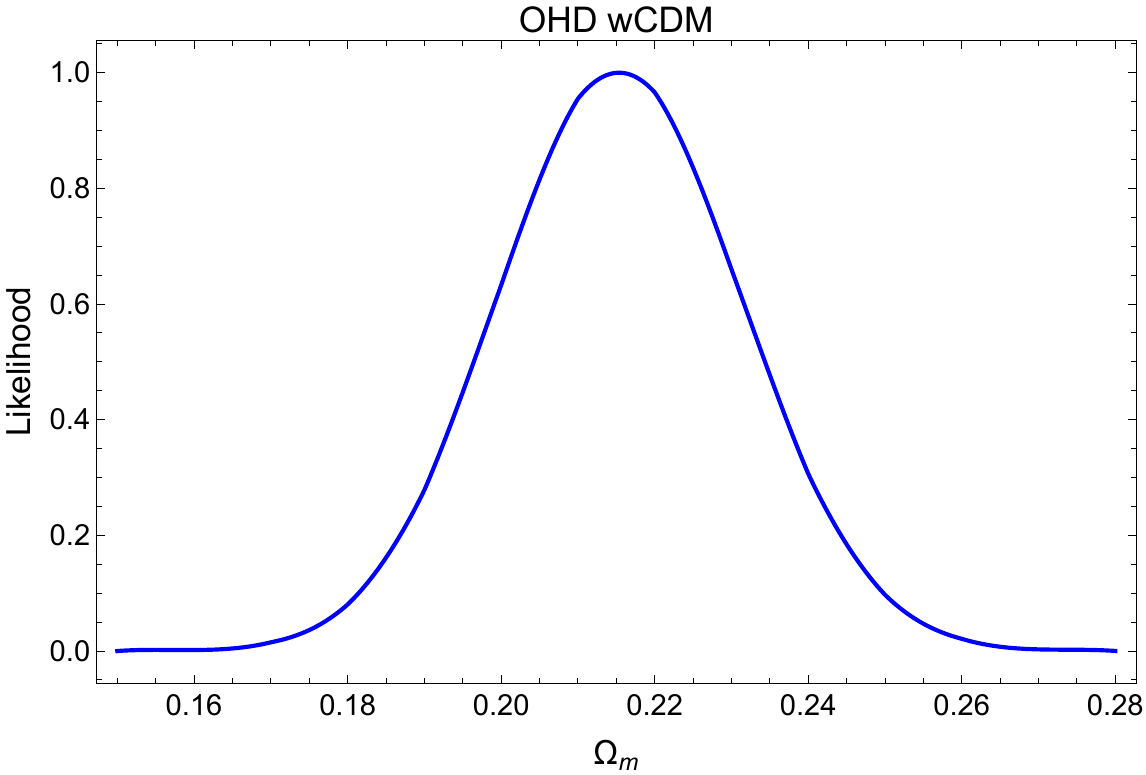}\\
            \includegraphics[width=2in,height=1in]{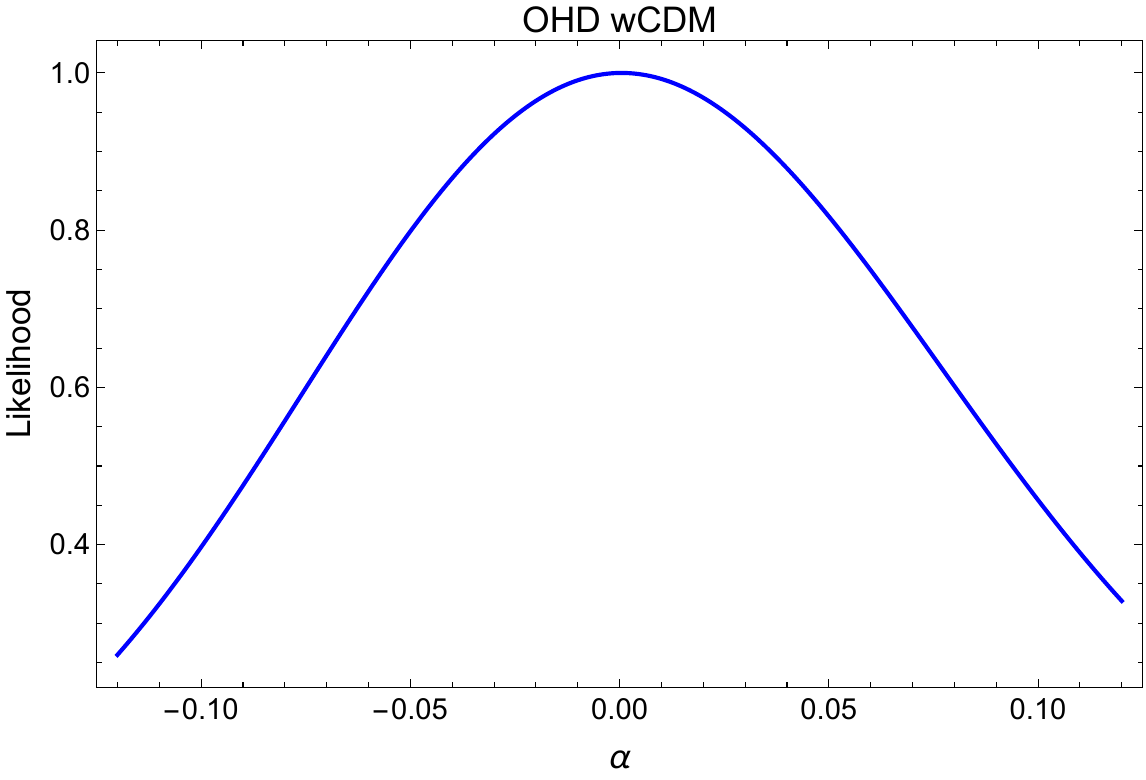}%
        } \\
    \end{tabular}
    \caption{\small\emph{$\Omega_m$ vs $\alpha$ graph with liklihood}} 
    \label{Comega}
\end{figure}
From the contour plot shown in Fig.~\ref{Comega}, the values of the parameters
$\alpha$ and $\Omega_m$ are determined for
$H_0 = 72~\mathrm{km\,s^{-1}\,Mpc^{-1}}$~\cite{sah} and $A = 0.034$.
The corresponding $1\sigma$ confidence intervals, obtained using the Hubble dataset,
are summarized in Table~\ref{t1}.
The Modified Chaplygin Gas (MCG) model has been extensively investigated in the literature,
and several studies have constrained its parameters using diverse observational datasets.
For instance, L.~Xu \textit{et al.}~\cite{lx} reported $\alpha = 0.000727^{+0.00142+0.00393}_{-0.00140-0.00234}$ and $A = 0.000777^{+0.000201+0.000915}_{-0.000302-0.000234}$ for $H_0 = 72.561^{+1.679+3.361}_{-1.690-3.223}$, which are comparatively smaller than the values obtained in the present analysis. Similarly, H.~Li \textit{et al.}~\cite{hl} derived parameter constraints from various data combinations; for example, using CMB+CC data they found $\alpha = -0.1688^{+0.1456+0.3340}_{-0.2143-0.3080}$ with $H_0 = 65.75^{+2.60+6.18}_{-3.71-5.13}$, whereas the combination CMB+JLA+CC yielded $\alpha = 0.0181^{+0.1029+0.3953}_{-0.2199-0.3060}$ with $H_0 = 68.07^{+2.72+5.30}_{-3.10-5.22}$.  On the other hand, J.~Lu \textit{et al.}~\cite{jl} obtained $\alpha = 1.724$ and $A = -0.085$, which are not physically viable within the MCG framework, since maintaining a subluminal sound speed requires $0 < \alpha < 1$ and $0 < A < \tfrac{1}{3}$. As pointed out earlier, Fabris \textit{et al.}~\cite{fab} ruled out the case of constant $A$, while Lima \textit{et al.}~\cite{lima} showed that a small value of $A$ is still admissible for $0.9 < \alpha < 1$.

Overall, although some discrepancies arise due to differences in datasets, priors, and statistical techniques, the general agreement among these studies underscores the robustness of the MCG model in describing the late-time accelerated expansion of the universe.

\begin{table}[h!]
\centering
\caption{Best-fit values of the $\Omega_m$ and $\alpha$ using Hubble-$57$ data (1$\sigma$ confidence region).}
\resizebox{\textwidth}{!}{%
\begin{tabular}{lccccc}
\hline
\textbf{Parameter} & $H_0~(\mathrm{km\,s^{-1}\,Mpc^{-1}})$ & $A$ & $\chi^2_m$ & $\Omega_m$ & $\alpha$ \\
\hline
\textbf{Best-fit values} & 72 & 0.034 & 46.48 & 0.216 & 0.0026 \\
\textbf{Range (1$\sigma$ region)} & -- & -- & -- & 0.1920, 0.2401 & $-0.1095, 0.1246$ \\
\hline
\end{tabular}
}
\label{t1}
\end{table}

We consider only positive values of $\alpha$ and $A$, as these parameters play crucial roles in defining the equation of state within the framework of the Modified Chaplygin Gas (MCG) model. Our analysis reveals that $\alpha$ assumes small positive values satisfying $0 < \alpha < 1$, consistent with previous results~\cite{dp25}. Similarly, we adopt small positive values of $A$, which favor the existence of the MCG model—contrary to the conclusions of Fabris et al.~\cite{fab}, who ruled out its viability. These parameter choices are well supported by current observational constraints, which strongly favor small positive values of both $\alpha$ and $A$ to ensure compatibility with cosmological data.

It is noteworthy that this restriction differs significantly from the pure Chaplygin Gas model, characterized by $\alpha = 1$ and $A = 0$. In contrast, the MCG model with $0 < \alpha < 1$ and $0 < A < \tfrac{1}{3}$ provides greater flexibility, enabling a smooth and realistic transition from a radiation-dominated era to a dark energy–dominated phase, thereby offering an improved fit to the observed evolution of the universe.

Now the present age of the universe is given by
\begin{equation}\label{eq:21}
t_0 = \int_{0}^{\infty} \frac{1}{(1+z)H(z)} dz.
\end{equation}
Using the parameter values listed in Table~\ref{t1} and substituting them into Eq.~\eqref{eq:21}, the present age of the Universe is found to be $t_0=13.79$ Gyr.  The value obtained is extremely close to the age inferred from the \textit{Planck} 2020 observations, which estimate the age of the Universe to be $13.8$ Gyr ~\cite{ros}. Therefore, the model parameters employed here lead to a cosmological history that is fully consistent with current high-precision measurements, reinforcing the viability of the framework in explaining the expansion age of the Universe.

\begin{figure}[h!]
    \centering 
    \begin{tabular}{l l}
        \parbox{2in}{%
            \includegraphics[width=2in,height=2 in]{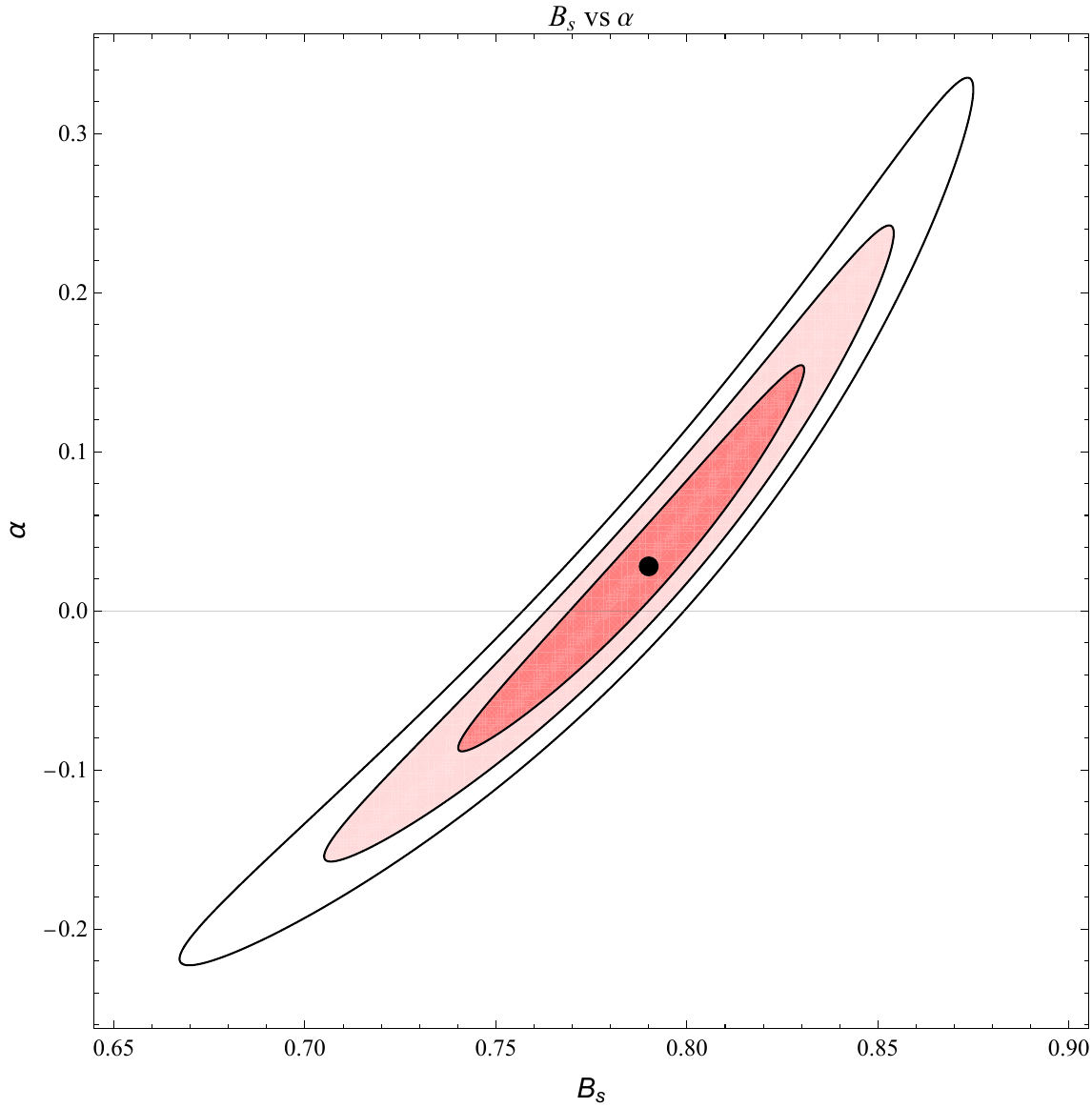}%

        } &
        \parbox{2in}{%
           \includegraphics[width=2in,height=1in]{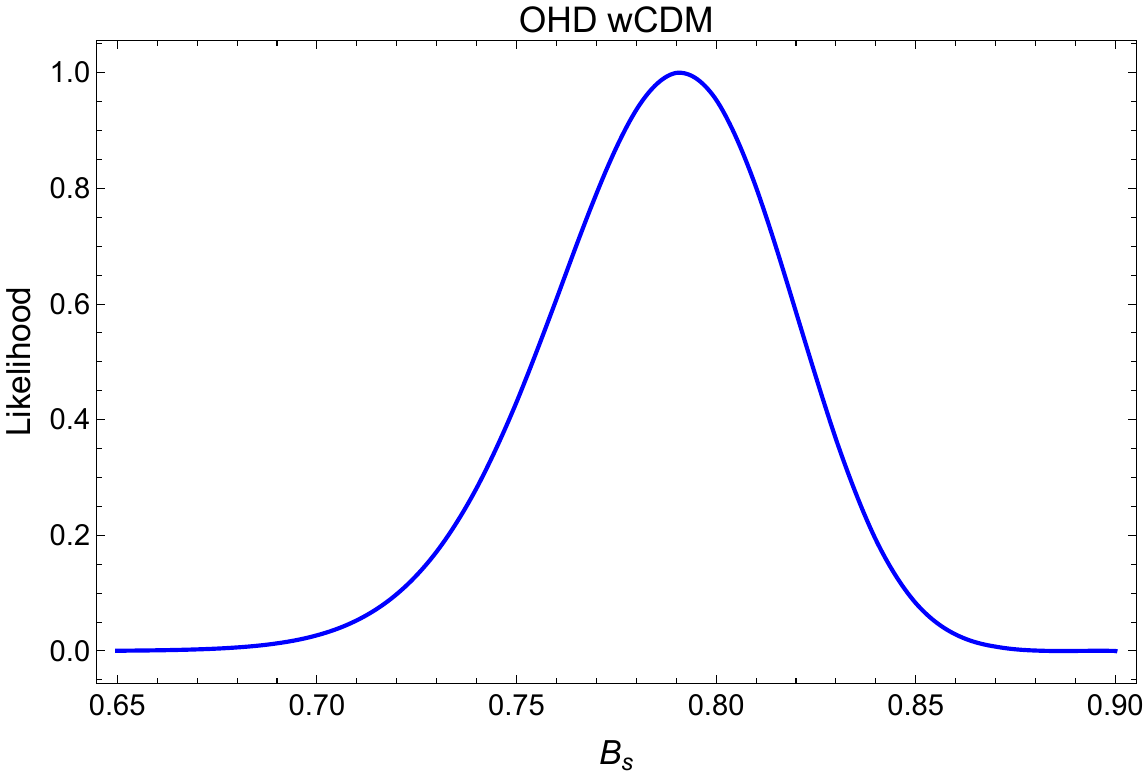}\\
            \includegraphics[width=2in,height=1in]{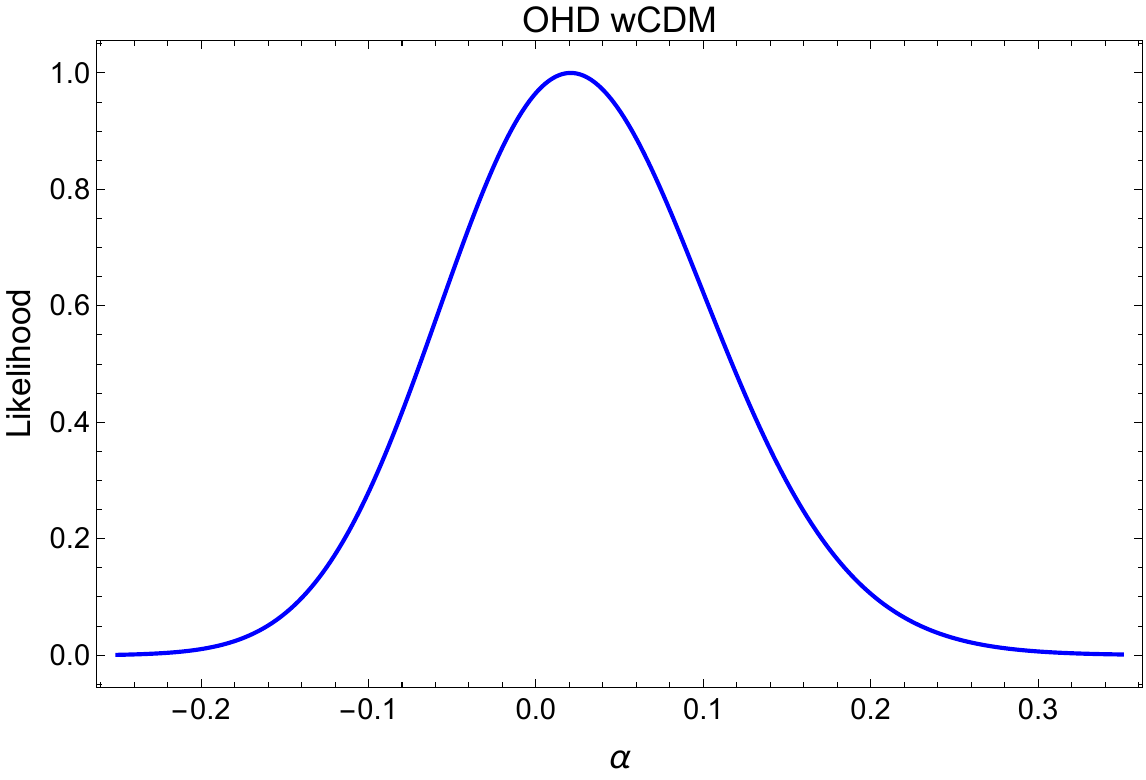}%
        } \\
    \end{tabular}
    \caption{\small\emph{$B_s$ vs $\alpha$ graph with liklihood}} 
    \label{bl}
\end{figure}

To constrain the model parameters we define $B_s = \left({\frac{B}{1+A}}\right)^{\frac{1}{1+\alpha}}\frac{1}{\rho_0}$ and  $c = (1-B_s^{1+\alpha})\rho_0^{1+\alpha}$. Using these definitions in Eq.~\eqref{eq:7}, the Hubble parameter can be written in terms of $B_s$ as

\begin{equation}\label{eq:21a}
H = H_0 \left \{ B_s^{1+\alpha}  +\left(1- B_s^{1+\alpha} \right)\left(1+z \right)^{3(1+A)(1+\alpha)} \right\}^{\frac{1}{2(1+\alpha)}}.
\end{equation}
The values of $B_s$ and $\alpha$ corresponding to the minimum $\chi^2$ are determined
from the contour plot shown in Fig.~\ref{bl},
for $H_0 = 72~\mathrm{km\,s^{-1}\,Mpc^{-1}}$ and $A = 0.025$.
The corresponding parameter ranges within the $1\sigma$ confidence region
are listed in Table~\ref{t2}.

\begin{table}[h!]
\centering
\caption{Best-fit values of the $B_s$ and $\alpha$ using Hubble-$57$ data (1$\sigma$ confidence region).}
\resizebox{\textwidth}{!}{%
\begin{tabular}{lccccc}
\hline
\textbf{Parameter} & $H_0~(\mathrm{km\,s^{-1}\,Mpc^{-1}})$ & $A$ & $\chi^2_m$ & $B_s$ & $\alpha$ \\
\hline
\textbf{Best-fit values} & 72 & 0.025 & 46.35 & 0.79 & 0.028 \\
\textbf{Range (1$\sigma$ region)} & -- & -- & -- & 0.7401, 0.8307 & $-0.0882, 0.1544$ \\
\hline
\end{tabular}
}
\label{t2}
\end{table}

It is worth noting that the earlier study by B. C. Paul et al. \cite{bcp} found  $A = 0.0101$, $B_s = 0.8450$ and $\alpha = 0.3403$, values that are marginally higher than those derived in the present analysis.

Using the parameter values in Table~\ref{t2} we obtain from Eqs.~\eqref{eq:21} and \eqref{eq:21a} $t_0 = 14.054$ Gyr. This estimate is marginally larger than the age reported by the \textit{Planck} 2020 analysis~\cite{ros}, but the difference lies within current observational uncertainties. Therefore, the result supports the consistency of the Modified Chaplygin Gas model with contemporary cosmological observations.

\subsubsection{Constraints from CMB Shift Parameter and Hubble-$57$ Data}
The Cosmic Microwave Background (CMB) provides a powerful probe of the early universe. A convenient and widely used quantity derived from CMB observations is the \textit{shift parameter}, which encodes the position of the first acoustic peak in the CMB power spectrum. The theoretical CMB shift parameter is defined as
\begin{equation}\label{eq:21b}
R_{\mathrm{th}} = \sqrt{\Omega_{m0}} \int_{0}^{z_*} \frac{dz}{E(z)},
\end{equation}
where $\Omega_{m0}$ is the present matter density parameter and $z_* \simeq 1090$ is the redshift of the last scattering surface. The observed value of the shift parameter is taken as \cite{na} $ R_{\mathrm{obs}} = 1.744 \pm 0.011$.
The corresponding chi-square function is defined as
\begin{equation}\label{eq:21c}
\chi^2_{\mathrm{CMB}} = \frac{\left(R_{\mathrm{th}} - R_{\mathrm{obs}}\right)^2}{\sigma_R^2},
\end{equation}
where $\sigma_R  (= 0.011) $ denotes the observational uncertainty.

\noindent Using Eq.~\eqref{eq:21a}, we evaluate the CMB shift parameter for the representative values $A=0.002$, $\alpha=0.002$, $\Omega_m = 0.2915$, and $B_s=0.709$. The analysis yields $R_{\mathrm{th}} = 1.734$, which is in very good agreement with the observed CMB value. The corresponding chi-square value is found to be $\chi^2_{\mathrm{CMB}} = 0.9066$, which lies well within the $1\sigma$ confidence level. This agreement indicates that the model successfully reproduces the background expansion history up to the recombination epoch ($z_* \sim 1090$). Therefore, the obtained result confirms the consistency of the Modified Chaplygin Gas model with early-universe observational constraints.

\noindent To obtain more stringent constraints, we carry out a combined analysis using the Hubble-$57$ dataset together with the CMB shift parameter. In this analysis, the parameters $A = 0.002$, $\alpha = 0.002$, and $\Omega_m = 0.2915$ are fixed, whereas $B_s$ and $H_0$ are allowed to vary freely. The best-fit values are determined through minimization of the total chi-square function, $\chi^2_{\mathrm{total}} = \chi^2_H + \chi^2_{\mathrm{CMB}}$, which yields  $\chi^2_{\mathrm{min}} = 21.66$, with the corresponding best-fit values $B_s = 0.695$ and
$H_0 = 69.41 \, \mathrm{km\,s^{-1}\,Mpc^{-1}}$.

\begin{figure}[h!]
    \centering 
    \begin{tabular}{l l}
        \parbox{2in}{%
            \includegraphics[width=2in,height=2in]{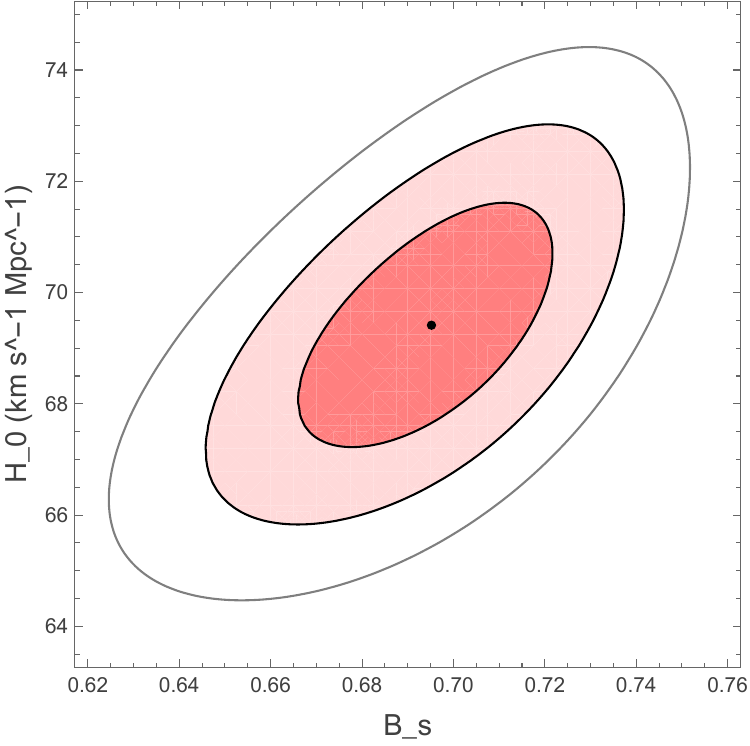}%

        } &
        \parbox{2in}{%
           \includegraphics[width=2in,height=1in]{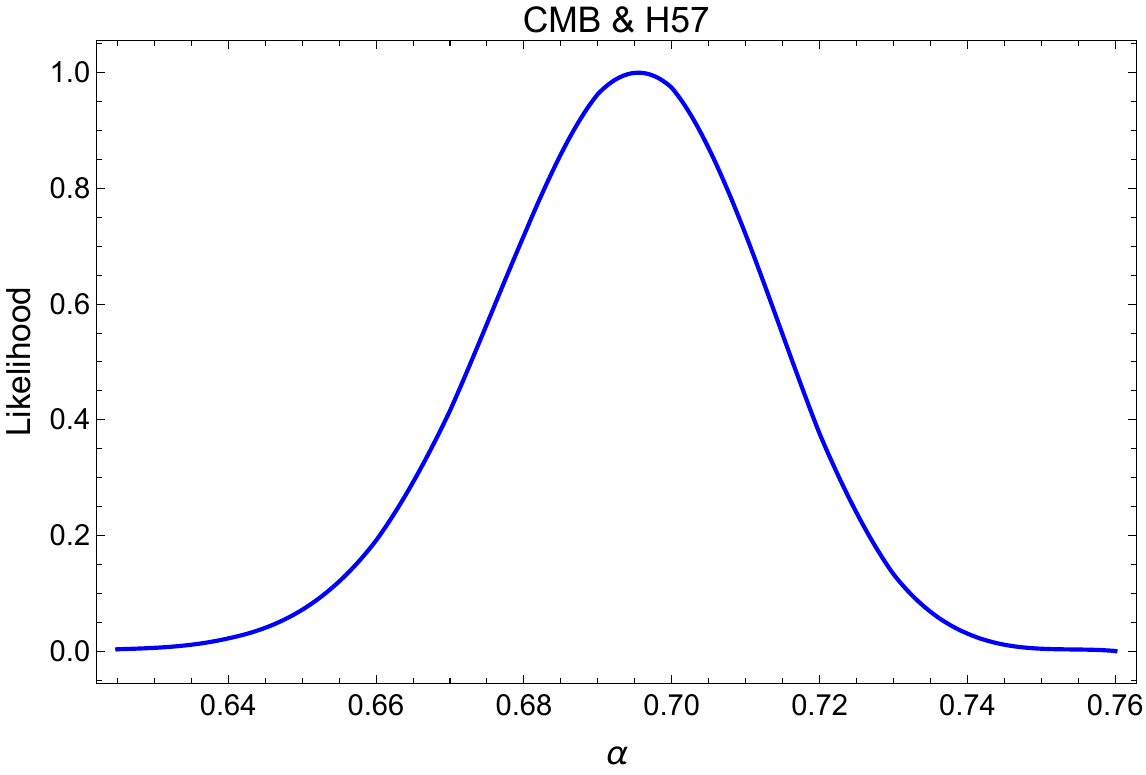}\\
            \includegraphics[width=2in,height=1in]{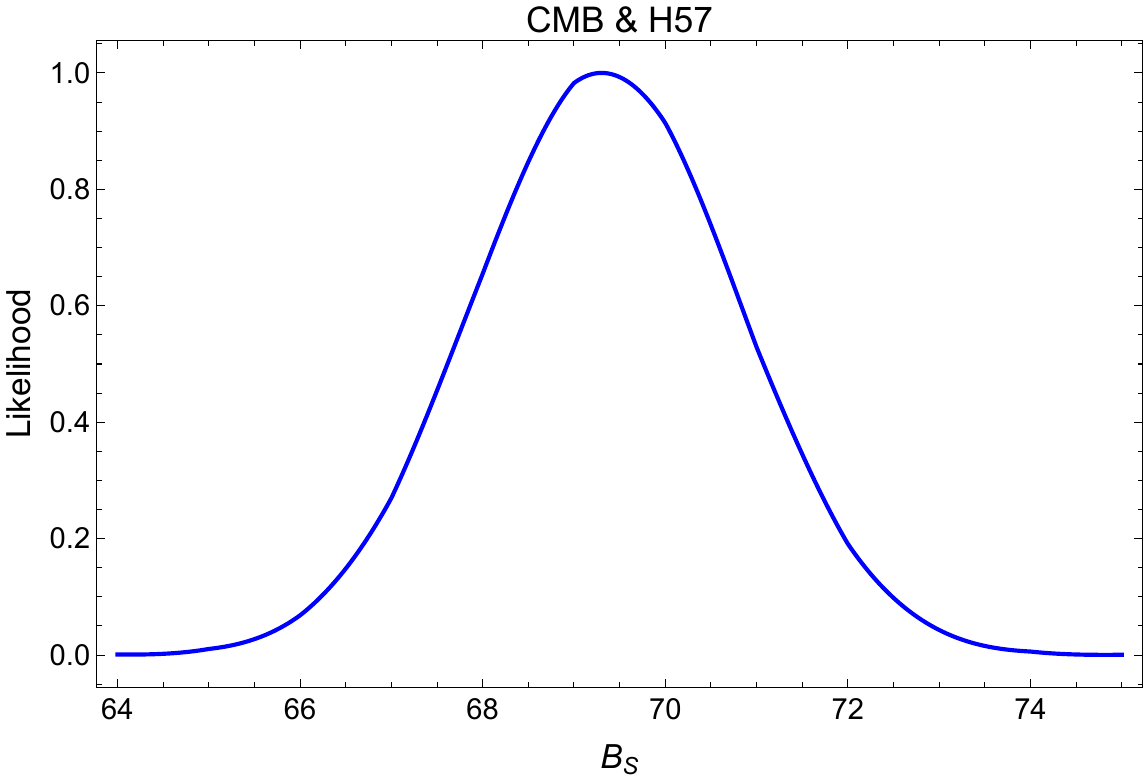}%
        } \\
    \end{tabular}
    \caption{\small\emph{$B_s$ vs $H_0$ graph with liklihood}} 
    \label{mch}
\end{figure}

\noindent The corresponding $1\sigma$ confidence intervals are determined from the contour analysis using $\Delta \chi^2 = 2.30$ for two parameters, resulting in $ B_s = 0.6952^{+0.0265}_{-0.0292}, H_0 = 69.41^{+2.20}_{-2.91} \mathrm{km\,s^{-1}\,Mpc^{-1}}$.  These results demonstrate that the inclusion of the CMB constraint significantly tightens the parameter space by incorporating early-universe information ($z_* \sim 1090$), thereby reducing degeneracies between parameters. The joint contour plots (Fig.~\ref{mch}) illustrate the allowed regions at the $1\sigma$, $2\sigma$, and $3\sigma$ confidence levels. Overall, the combined Hubble-$57$ and CMB analysis provides a coherent and observationally consistent description of cosmic evolution across both early and late epochs.

\subsection{Possibilities of Structure Formation }

A key requirement for any viable cosmological model is the existence of a sufficiently extended matter-dominated phase, during which large-scale structures can form through gravitational instability. In the present Modified Chaplygin Gas (MCG) model, the effective equation-of-state parameter evolves smoothly from   early universe stage to a dark energy dominated phase at late times.

In the high-density regime, the effective equation-of-state parameter reduces to $w_{\mathrm{eff}} \approx A$. For the best-fit values obtained in this work, $A \sim 0.002$, which is extremely small. Consequently, the effective pressure remains negligible over a substantial period of cosmic evolution, leading to a reasonably prolonged quasi-matter-dominated phase. As illustrated in Fig.~\ref{wz2}, $w_{\mathrm{eff}}$ remains very close to zero throughout the redshift interval $3.5 \leq z \leq 12 $, indicating an extended matter-like epoch rather than a rapid transition. During this period, the energy density evolves approximately as $\rho \propto a^{-3}$, closely mimicking the standard cold dark matter behavior. Such a sufficiently prolonged matter-like phase provides favorable conditions for the growth of density perturbations and the formation of large-scale structures before the onset of the late-time accelerated expansion.

At the perturbative level, the growth of matter density fluctuations is governed by the standard evolution equation~\cite{ya}
\begin{equation}\label{eq:21d }
\ddot{D} + 2H \dot{D} - 4\pi G \rho_{\mathrm{eff}} D = 0,
\end{equation}
which, in the quasi-matter-dominated phase ($w_{\mathrm{eff}} \approx 0$), reduces to the standard form admitting the growing mode $D(a) \propto a$. This demonstrates that the model supports efficient growth of density perturbations, consistent with large-scale structure formation.

Furthermore, the combined Hubble-$57$ and CMB analysis yields $\Omega_m \sim 0.3$ and shows excellent agreement with the CMB shift parameter ($z_* \sim 1090$), ensuring consistency with both early- and late-time observations.

Therefore, the smallness of the parameter $A$ not only guarantees a prolonged matter-like phase but also naturally ensures that the MCG model remains compatible with the requirements of structure formation, while simultaneously accounting for the late-time accelerated expansion of the universe.

As discussed earlier, the key Eq.~\eqref{eq:8} does not admit an explicit analytical solution in a simple functional form of time. Consequently, it is not possible to express the evolution of cosmological quantities—such as the scale factor, the flip time, and other dynamical parameters—in a closed form. To overcome this limitation and to determine the flip time along with other physical characteristics of the cosmological model, we adopt an \emph{ alternative approximation method}~\cite{dp25, dp19} in the following section.

\vspace{0.2 cm}
\section{An Alternative Formulation }
Since the scale factor increases with time in an expanding universe, the last term in Eq.~\eqref{eq:8} becomes negligible compared to the second term when expanded binomially. Therefore, the results obtained from a first-order approximation of the expanded Eq.~\eqref{eq:8} are particularly relevant. In what follows, we present the exact solution corresponding to this first-order approximation. Accordingly, at the late stage of cosmic evolution, the first-order approximation of Eq.~\eqref{eq:8} takes the form
\begin{equation}\label{eq:22}
3 \frac{\dot{a}^{2}}{a^{2}}=
\left(\frac{B}{1+A}\right)^{\frac{1}{1+\alpha}} +
\frac{1}{1+\alpha}
\left(\frac{1+A}{B}\right)^{\frac{\alpha}{1+\alpha}}~\frac{c}{a^{3(1+A)(1+\alpha)}}.
\end{equation}
For economy of space we skip the intermediate steps and write the
final solution as,

\begin{equation}\label{eq:23}
 a(t) = a_{0} \mathrm{sinh}^{n}\omega t,
 \end{equation}
 where,
$ a_{0} = \left\{\frac{c (1+A)}{B (1+\alpha)}\right
\}^{\frac{1}{3(1+A)(1+\alpha)}}$,
 $n =  \frac{2}{3(1+A)(1+\alpha)}$ and $\omega =
\frac{\sqrt{3}}{2}(1+A)^{\frac{1+2\alpha}{2(1+\alpha)}}(1+\alpha)B^{\frac{1}{2(1+\alpha)}}$.

\begin{figure}[ht]
\begin{center}
    \includegraphics[width=7.2 cm]{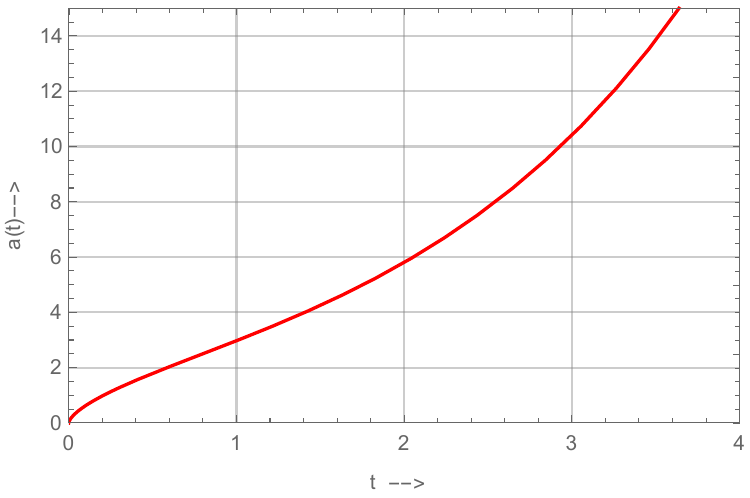}
      \caption{
  \small\emph{The variations of $a(t)$ and $t$. }\label{at}
    }
\end{center}
\end{figure}
The evolution of the scale factor  $a(t)$  with time $t$ for the Modified Chaplygin Gas (MCG) model is shown in Fig.~\ref{at}. While the MCG model is usually studied in its asymptotic limits, a first-order approximation reveals a continuous evolution of $a(t)$  from an early decelerating phase to a late-time accelerated expansion, as also indicated by Eq.~\eqref{eq:23}.

Using Eqs.~\eqref{eq:3}, \eqref{eq:4}   and \eqref{eq:23} we can write the expression of density and  pressure as follows,
\begin{equation}\label{eq:24}
 \rho = 3n^2 \omega^2 \coth^2{\omega t} = \left(\frac{B}{1+A}\right)^{\frac{1}{1+\alpha}} \left \{1 + (1+z)^{\frac{2}{n}} \right\},
\end{equation}
and
\begin{equation}\label{eq:25}
 p = n\omega^2 \left \{(2-3n)\coth^2\omega t -2 \right\}
  = - \left(\frac{B}{1+A}\right)^{\frac{1}{1+\alpha}} \left [1 - \{\alpha + A(1+ \alpha) \} (1+z)^{\frac{2}{n}} \right].
\end{equation}
We now discuss the evolution of various cosmological parameters within this first-order approximation framework.

\subsection{\textbf{  Deceleration Parameter :}}
From Eq.~\eqref{eq:23}, we obtain  the expression of deceleration parameter as
\begin{equation}\label{eq:26}
q = - \frac{1}{H^{2}} \frac{\ddot{a}}{a}
= \frac{1-n \, \cosh^{2}(\omega t)}{n \, \cosh^{2}(\omega t)}
= \frac{1}{n \left\{ 1 + (1+z)^{-\tfrac{2}{n}} \right\}} - 1.
\end{equation}
This expression shows that the exponent $n$ determines the  nature of evolution of the deceleration parameter $q$. A closer inspection reveals that
\begin{figure}[ht]
\begin{center}
    \includegraphics[width=7.2 cm]{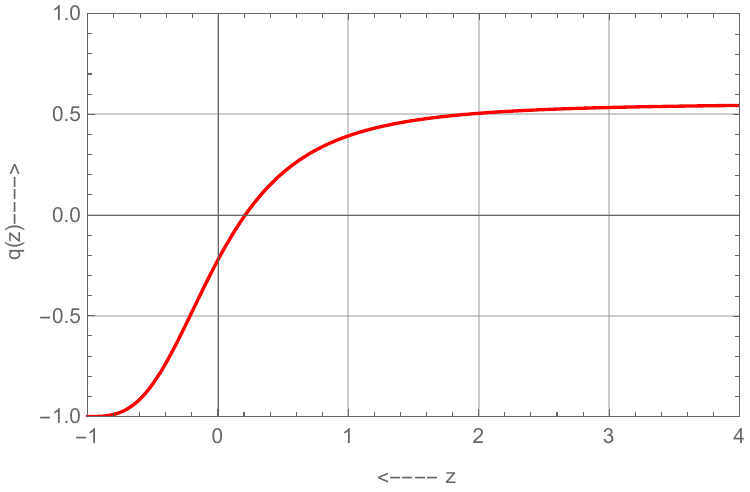}
      \caption{
  \small\emph{The variations of $ q (z)$ and $z$. }\label{qzaa}
    }
\end{center}
\end{figure}
\begin{enumerate}[(i)]
    \item $n \geq 1$ gives always acceleration, no flip occurs in this condition. But for $n \geq 1$, $\alpha < - \frac{1+3A}{3(1+A)}$ which is physically unrealistic for $A > 0$ \& $\alpha > 0$.

\item  $0< n < \frac{2}{3}$ gives the early deceleration and late acceleration in the conditions for $A > 0$ \& $\alpha > 0$, so desirable feature of \emph{flip} occurs which agrees with the observational analysis.
\end{enumerate}
 At the time of flip $q = 0$ and the flip time  $t_f$ is given by

\begin{equation}\label{eq:27}
t_{f}  = \frac{1}{\omega} \cosh^{-1} \left(\frac{1}{\sqrt{n}}
\right) = \frac{1}{\omega} \cosh^{-1}\sqrt{\frac{3(1+A)(1+\alpha)}{2}}.
\end{equation}

\begin{figure}[ht]
\begin{center}
    \includegraphics[width=7.2 cm]{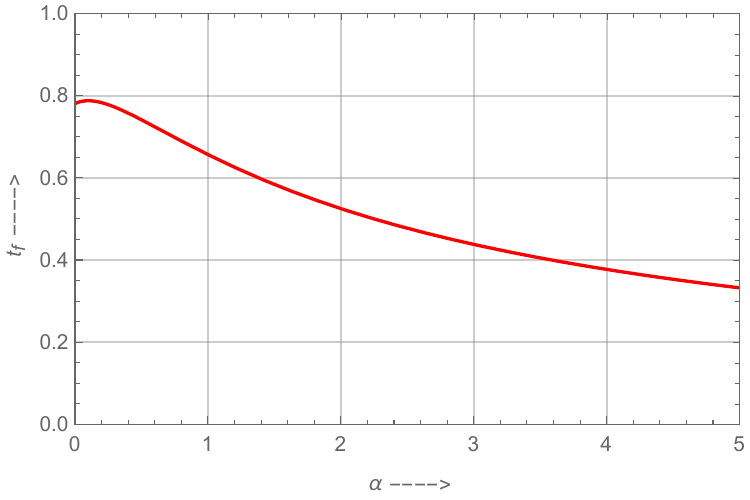}
      \caption{
  \small\emph{The variations of $ t_f$ and $\alpha$. }\label{tfa}
    }
\end{center}
\end{figure}
Fig.~\ref{tfa} shows the flip time $t_f$  as a function of
$\alpha$. The plot clearly indicates that smaller values of
$\alpha$  correspond to a later transition to the accelerated phase of expansion. This behavior is consistent with Eq.~\eqref{eq:1b} for larger  $\alpha$, the negative pressure associated with the fluid grows more rapidly with cosmic expansion, causing the Universe to enter the accelerating regime earlier. Therefore, the trend observed in Fig.~\ref{tfa} is fully consistent with the theoretical expectations derived from Eq.~\eqref{eq:1b}.

Now the redshift at the time of flip is
\begin{equation}\label{eq:27a}
z_{f}  = \left(\frac{n}{1-n}\right)^{\frac{n}{2}} - 1.
\end{equation}
To obtain acceleration at the present epoch, the condition $z_f > 0$ must hold, which leads to $\alpha < \frac{1-3A}{3(1+A)} $. This further implies that $A < \tfrac{1}{3}$ in order to ensure $\alpha > 0$, indicating that the model parameters must lie within these bounds to produce a physically consistent universe dominated by dark energy–like behavior, characterized by negative effective pressure and accelerated expansion.

Now we discuss the evolution of deceleration parameter at different epoch as follows:
\begin{enumerate}[(i)]
    \item At the early epoch of the universe, corresponding to high redshift $z$, we obtain from Eq.~\eqref{eq:26}
\begin{equation}\label{eq:28}
q = \frac{1}{n} - 1 = \frac{3(1+A)(1+\alpha)}{2} - 1.
\end{equation}
For $q$ to be positive, the condition
$ \alpha > - \frac{1+3A}{3(1+A)} $
must be satisfied. In particular, for any positive value of $A$, this implies that $\alpha$ must also be positive.

\item At the present epoch, i.e., at $z=0$, Eq.~\eqref{eq:26} reduces to
\begin{equation}\label{eq:29}
q = \frac{1}{2n} - 1 = \frac{3(1+A)(1+\alpha)}{4} - 1.
\end{equation}
Since the present universe is undergoing accelerated expansion, we require $q<0$. This condition implies  $ \alpha < \frac{4}{3(1+A)} - 1$.
Moreover, using the bound $0 \leq A \leq \tfrac{1}{3}$, we conclude that $ A < \tfrac{1}{3}$.

\item At the late universe, the Eq.~\eqref{eq:26} gives $q = -1$ which corresponds to $\Lambda$CDM model.

\end{enumerate}

Fig.~\ref{qzaa} shows the deceleration parameter $q$ versus redshift $z$ in the Modified Chaplygin Gas model, illustrating a smooth transition from early deceleration ($q>0$) to late-time acceleration ($q<0$) at the flip redshift $z_f$, without separate dark matter or dark energy components.

\subsection{\textbf{Effective equation of state:}}
The effective equation of state $w_e$ is given by
\begin{equation}\label{eq:30}
w_e = \frac{p}{\rho} = \frac{2-3n}{n^2} - \frac{2}{3} \tanh^2 \omega t = - \frac{1-\{\alpha+ A (1 +\alpha)\}(1+z)^{\frac{2}{n}}}{1+(1+z)^{\frac{2}{n}}}.
\end{equation}

\begin{figure}[ht]
    \centering
    \subfigure[\small\emph{$ -1 \leq z \leq 12 $}]{
        \includegraphics[width=6 cm]{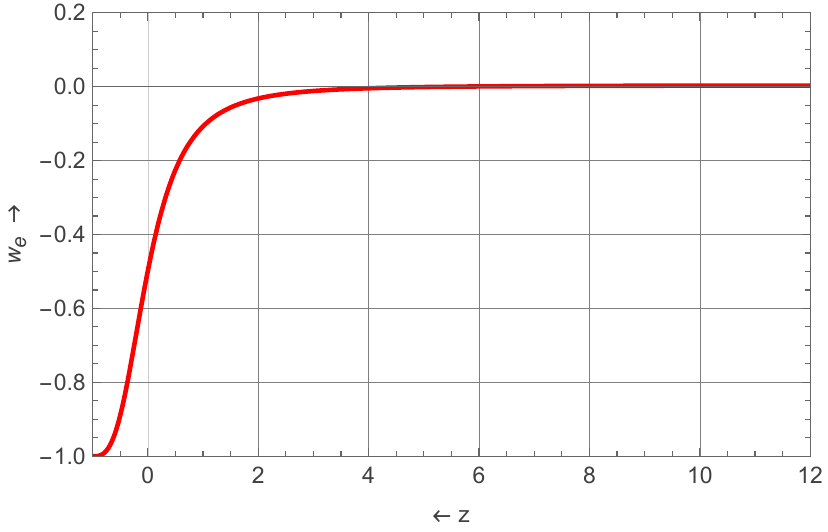}
        \label{wzaa1}
    }
    ~~~
    \subfigure[\small\emph{$2.5 \leq z \leq 12 $}]{
        \includegraphics[width=6 cm]{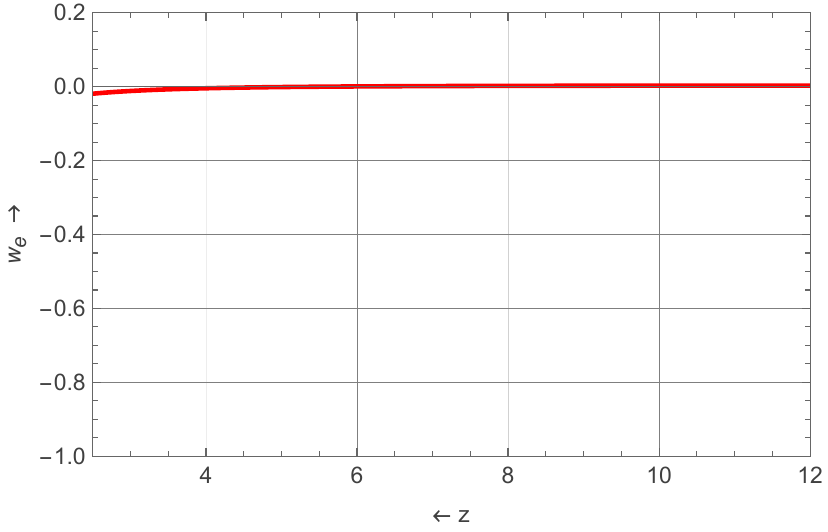}
        \label{wzaa2}
    }
    \caption[\small Optional caption for list of figures]{
        \small\emph{The variations of $w_{\text{e}}$ and $z$}
    }
    \label{wz}
\end{figure}

\begin{enumerate}[(i)]

\item Early Epoch: In the high-redshift limit, the effective equation-of-state parameter can be expressed, using Eq.~\eqref{eq:30}, as
\begin{equation}\label{eq:31}
w_e \approx \alpha + A(1+\alpha),
\end{equation}
which remains positive for all physically admissible positive values of $A$ and $\alpha$. Consequently, the model describes a decelerating phase in the early universe, characterized by a positive effective pressure.

\item Dust dominated Epoch: In the dust-dominated era, where $w_e = 0$, the corresponding redshift $z_{dust}$ can be obtained from Eq.~\eqref{eq:30} as
\begin{equation}\label{eq:32}
z_{dust} = \left\{ \frac{1}{\alpha + A(1+\alpha)} \right\}^{\frac{n}{2}} - 1.
\end{equation}
This relation shows that $ z_{dust} > z_f$, which is consistent with the expected cosmological timeline. It signifies that the transition to an accelerating phase  occurs after the dust-dominated epoch—marking the evolution from matter domination to a dark energy–driven universe.

\item Present Epoch: At $z = 0$, Eq.~\eqref{eq:30} reduces to
\begin{equation}\label{eq:33}
w_e \approx - \frac{1 - \left\{ \alpha + A(1+\alpha) \right\}}{2}.
\end{equation}
Since $\alpha < \tfrac{1-3A}{3(1+A)}$ and $A < \tfrac{1}{3}$, we have $w_e < 0$, indicating that the present universe is undergoing an accelerated expansion.

\item Late-Time Limit: As $z \to -1$, Eq.~\eqref{eq:30} reduces to  $w_e \approx -1$,
showing that the model asymptotically approaches a cosmological constant–dominated phase, effectively reproducing the $\Lambda$CDM scenario.

\end{enumerate}

Fig.~\ref{wzaa1} illustrates the evolution of the effective equation-of-state parameter $w_{\mathrm{e}}$ over the redshift range $-1 \leq z \leq 12$ in the approximation approach of the Modified Chaplygin Gas (MCG) model. At high redshifts, $w_{\mathrm{e}}>0$ corresponds to a decelerated matter-like phase, while near the present epoch, $-1<w_{\mathrm{e}}<0$ signals the onset of accelerated expansion driven by an effective dark-energy component. At late times, $w_{\mathrm{e}}\rightarrow -1$, indicating that the model asymptotically approaches the $\Lambda$CDM limit.

To further illustrate the intermediate-redshift behavior, Fig.~\ref{wzaa2} presents an enlarged view over the range $2.5 \leq z \leq 12$. It is evident that $w_{\mathrm{e}}$ remains very close to zero throughout this interval, reflecting the existence of an extended quasi-matter-dominated epoch. This behavior is a direct consequence of the small best-fit value of the parameter $A=0.0012$, which keeps the effective pressure negligible over a substantial period of cosmic evolution. As a result, the model closely mimics the standard matter-dominated era before evolving toward the late-time accelerated phase. Thus, the approximation approach reproduces the same qualitative behavior discussed in Sec.~3.3, namely a smooth transition from a prolonged matter-like epoch to the $\Lambda$CDM limit at late times.

Now, we aim to constrain the model parameters and examine the observational consistency of the present approach with the previously discussed MCG model in Sec.~3.4.

\subsection{ Observational constraint:}

Here, we first constrain the model parameters using the Hubble-$57$ dataset and subsequently perform a combined Hubble-$57$ \& CMB analysis to obtain more reliable constraints.

\subsubsection{Constraints from Hubble-$57$ dataset}

\begin{figure}[h!]
    \centering 
    \begin{tabular}{l l}
        \parbox{2in}{%
            \includegraphics[width=2in,height=2in]{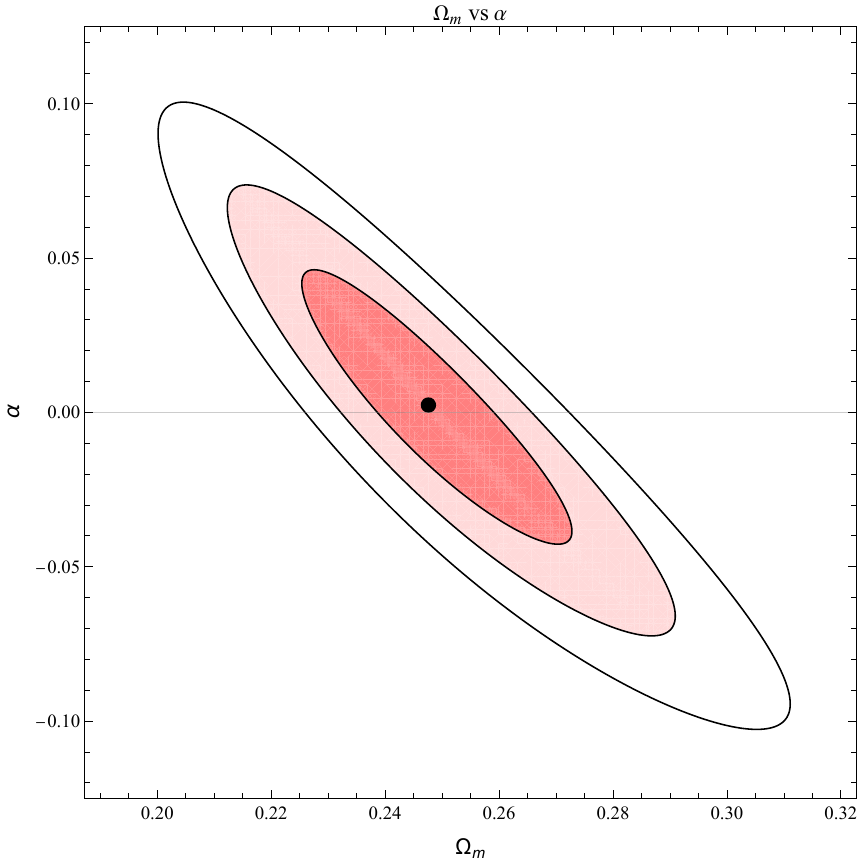}%

        } &
        \parbox{2in}{%
           \includegraphics[width=2in,height=1in]{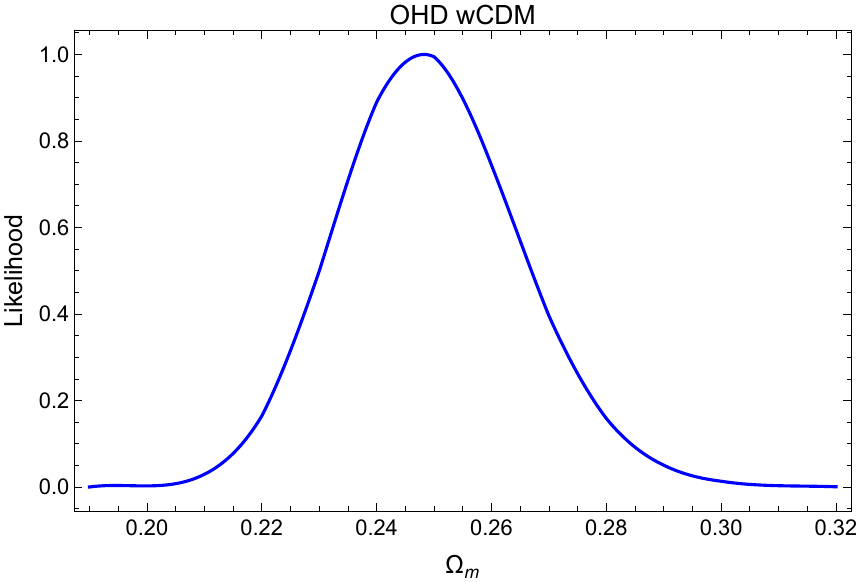}\\
            \includegraphics[width=2in,height=1in]{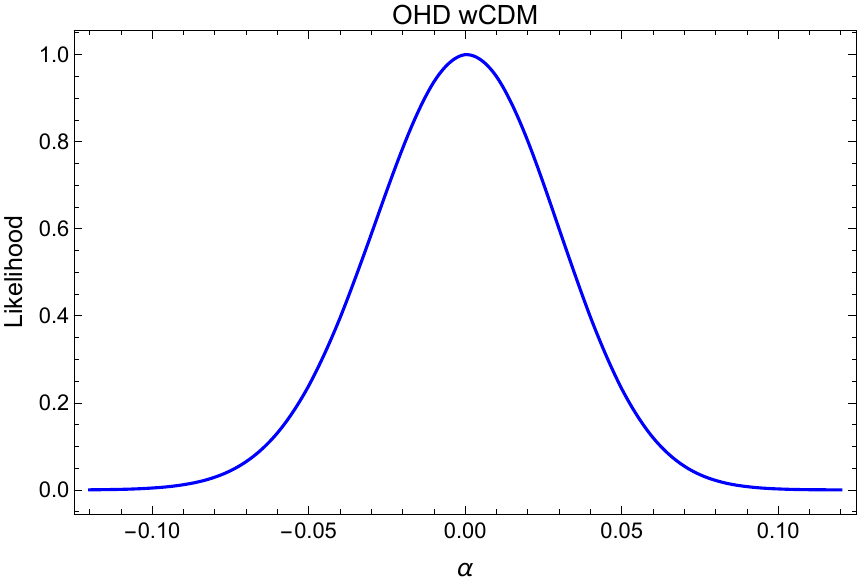}%
        } \\
    \end{tabular}
    \caption{\small\emph{$\Omega_m$ vs $\alpha$ graph with liklihood} }
    \label{Coaa1}
\end{figure}

Since the Modified Chaplygin Gas (MCG) model has been extensively studied and its parameters have been constrained by several authors using various observational datasets, it is important to verify the consistency of our results. As our solution is obtained through an approximation method, we examine whether the derived parameter values (see Sec.~3) are consistent with observational constraints. For this purpose, we consider the Hubble-$57$ dataset throughout the analysis.
The Hubble parameter can now be expressed in terms of the matter density parameter $\Omega_m \left(= \frac{c}{\rho_0^{1+\alpha}}\right)$ using Eq.~\eqref{eq:22} as
\begin{equation}\label{eq:34}
H = H_0 \Omega_m^{\frac{1}{1+\alpha}}\left \{\left(\frac{1 - \Omega_m}{\Omega_m}\right)^{\frac{1}{1+\alpha}}  + \frac{1}{1 + \alpha}\left(\frac{\Omega_m}{1 - \Omega_m}\right)^{\frac{\alpha}{1+\alpha}}
(1+z)^{3(1+A)(1+\alpha)}\right\}^{\frac{1}{2}}.
\end{equation}
A contour plot, shown in Fig.~\ref{Coaa1}, has been generated using Eq.~\eqref{eq:34}.
The best-fit values of the parameters $\alpha$ and $\Omega_m$ for
$H_0 = 71~\mathrm{km\,s^{-1}\,Mpc^{-1}}$ and $A = 0.002$,
along with their corresponding $1\sigma$ confidence ranges,
are listed in Table~\ref{t3}.

\begin{figure}[h!]
    \centering 
    \begin{tabular}{l l}
        \parbox{2in}{%
            \includegraphics[width=2in,height=2.1in]{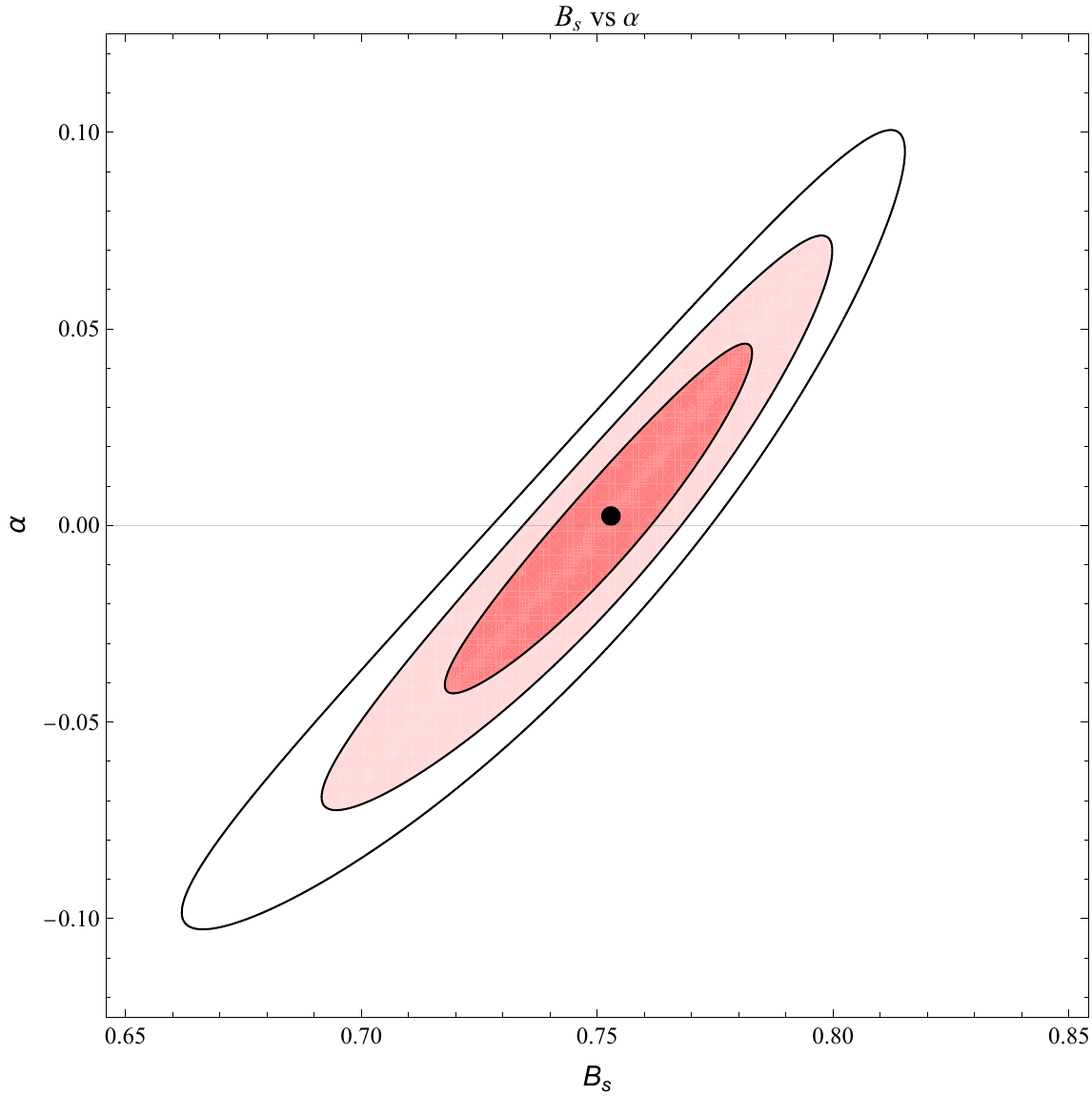}%

        } &
        \parbox{2in}{%
           \includegraphics[width=2in,height=1in]{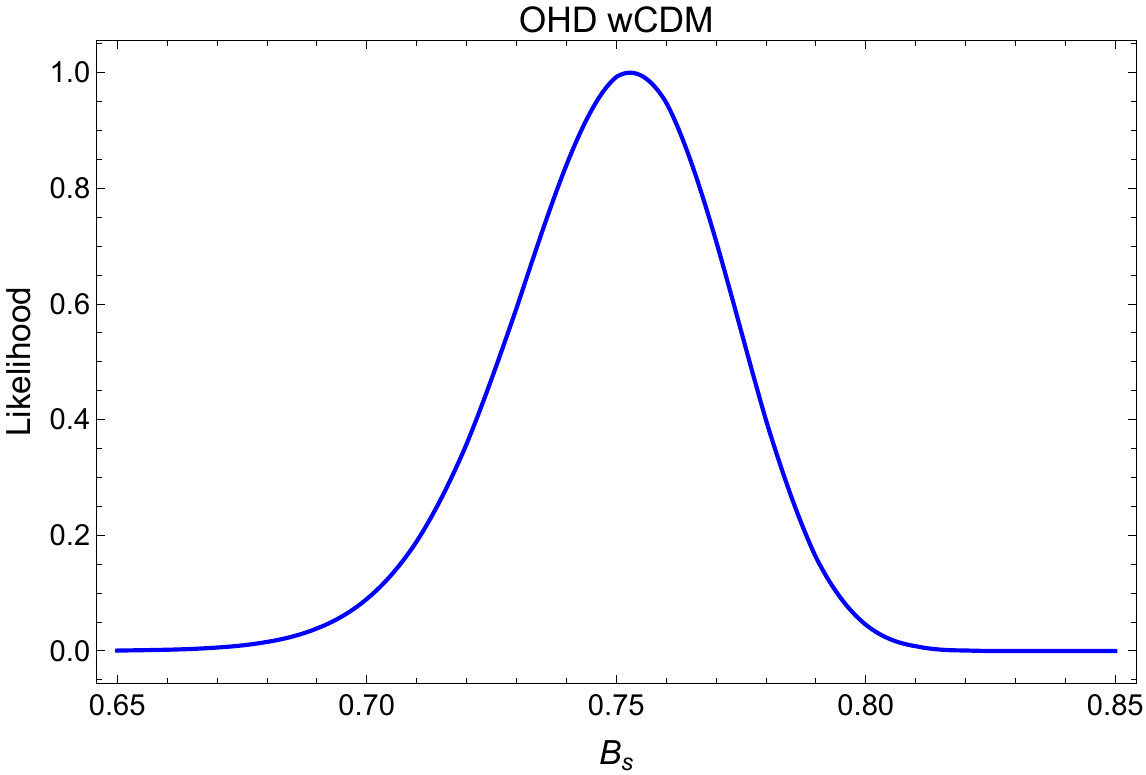}\\
            \includegraphics[width=2in,height=1in]{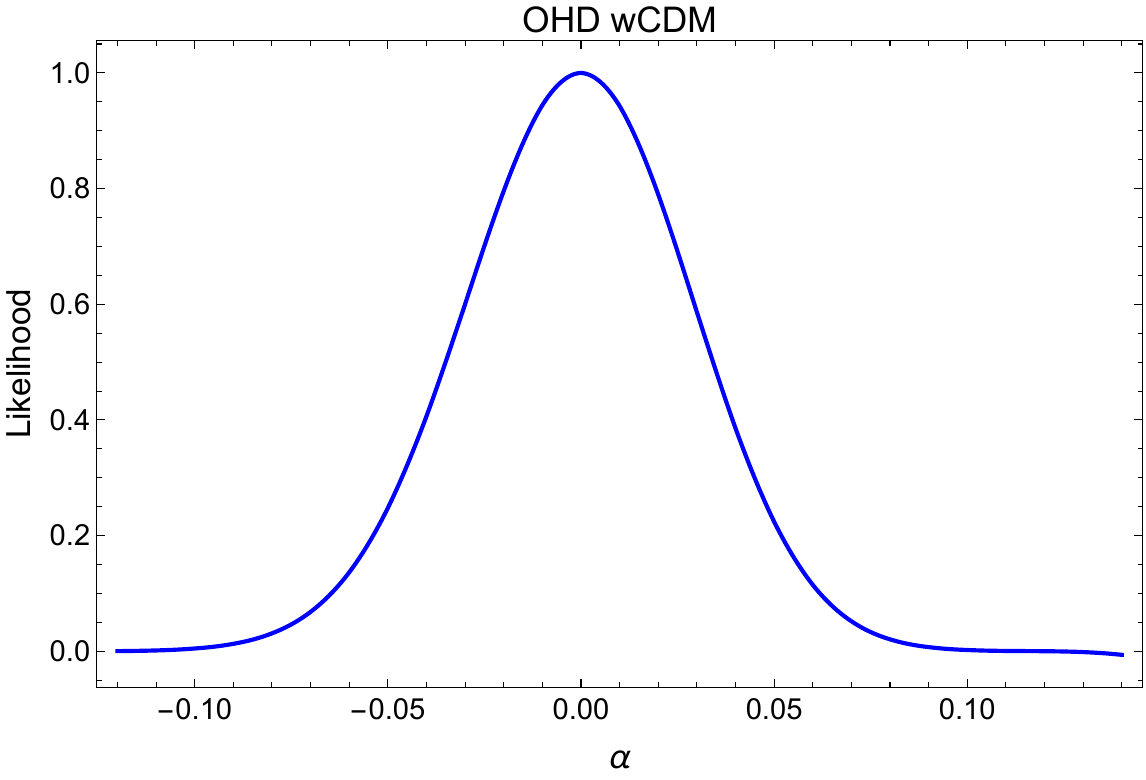}%
        } \\
    \end{tabular}
    \caption{\small\emph{$B_s$ vs $\alpha$ graph with liklihood}} 
    \label{Lbsa}
\end{figure}

\begin{table}[h!]
\centering
\caption{Best-fit values of the $\Omega_m$ and $\alpha$ using Hubble-$57$ data (1$\sigma$ confidence region).}
\resizebox{\textwidth}{!}{%
\begin{tabular}{lccccc}
\hline
\textbf{Parameter} & $H_0~(\mathrm{km\,s^{-1}\,Mpc^{-1}})$ & $A$ & $\chi^2_m$ & $\Omega_m$ & $\alpha$ \\
\hline
\textbf{Best-fit values} & 71 & 0.002 & 44.87 & 0.248 & 0.0024 \\
\textbf{Range (1$\sigma$ region)} & -- & -- & -- & $0.2254$, $0.2728$ & $-0.0427$,$ 0.0462$ \\
\hline
\end{tabular}
}
\label{t3}
\end{table}
From Eqs.~\eqref{eq:21} and \eqref{eq:34}, and using the parameter values listed in Table~\ref{t3}, we obtain the present age of the universe as $t_0 = 13.94$~Gyr. This value is slightly higher than the estimate inferred from the Planck 2020 results~\cite{ros}.
Furthermore, to constrain the model parameters, we define
$B_s = \left(\frac{B}{1+A}\right)^{\frac{1}{1+\alpha}}\frac{1}{\rho_0}$,
which allows the Hubble parameter to be expressed in terms of $B_s$ as
\begin{equation}\label{eq:35}
H = H_0 \left \{ B_s  +\frac{1}{1+\alpha}\frac{\left(1- B_s^{1+\alpha} \right)}{B_s^{\alpha}}\left(1+z \right)^{3(1+A)(1+\alpha)}
\right\}^{\frac{1}{2}}.
\end{equation}
The values of $B_s$ and $\alpha$ corresponding to the minimum $\chi^2_{m}$
are determined from the contour plot shown in Fig.~\ref{Lbsa},
for $H_0 = 71~\mathrm{km\,s^{-1}\,Mpc^{-1}}$ and $A = 0.002$,
with the corresponding $1\sigma$ confidence ranges presented in Table~\ref{t4}.
It is found that $\alpha < 1$, in contrast to the pure Chaplygin Gas model,
for which $\alpha = 1$.

It is worth mentioning that the condition derived in Sec.~3.2, $B > \frac{c}{2}(1+3A)(1+A)$, or equivalently $\frac{B_s^{1+\alpha}}{\Omega_m}> \frac{1+3A}{2}$, is also satisfied by the values obtained from our alternative approach, thereby confirming the consistency of the results.

\begin{table}[h!]
\centering
\caption{Best-fit values of the $B_s$ and $\alpha$ using Hubble-$57$ data (1$\sigma$ confidence region).}
\resizebox{\textwidth}{!}{%
\begin{tabular}{lccccc}
\hline
\textbf{Parameter} & $H_0~(\mathrm{km\,s^{-1}\,Mpc^{-1}})$ & $A$ & $\chi^2_m$ & $B_s$ & $\alpha$ \\
\hline
\textbf{Best-fit values} & 71 & 0.002 & 44.87 & 0.7527 & 0.0024 \\
\textbf{Range (1$\sigma$ region)} & -- & -- & -- & 0.7177, 0.7828 & $-0.0427, 0.0462$ \\
\hline
\end{tabular}
}
\label{t4}
\end{table}

We also determine the present age of the universe using the parameter values from Table~\ref{t4}. From Eqs.~\eqref{eq:21} and \eqref{eq:35}, we calculate $t_0 = 13.94$ Gyr.  This value is slightly higher than the result obtained from Planck 2020 data~\cite{ros}. This slight discrepancy may be attributed to the fact that these datasets probe different epochs and aspects of cosmic evolution. Moreover, variations in the assumed values of the Hubble constant and other cosmological parameters across different analyses naturally lead to minor differences in the estimated age of the universe.\\

\subsubsection{Constraints from CMB Shift Parameter and Hubble-$57$ Data}
\noindent Using Eq.~\eqref{eq:35} with the chosen  values $A=0.0012$, $\alpha=0.0024$, $\Omega_m = 0.301$, and $B_s=0.70$, we obtain $R_{\mathrm{th}} = 1.734$ with $\chi^2_{\mathrm{CMB}} = 0.909$, which agrees well with observations within the $1\sigma$ confidence level and confirms consistency with early-universe constraints.

\noindent To obtain tighter constraints, we perform a joint Hubble-$57$ + CMB analysis by fixing $A = 0.0012$, $\alpha = 0.0024$, and $\Omega_m = 0.301$, while treating $B_s$ and $H_0$ as free parameters. Minimization of $\chi^2_{\mathrm{total}} = \chi^2_H + \chi^2_{\mathrm{CMB}}$ yields the best-fit values
$\chi^2_{\mathrm{min}} = 21.93$, $B_s = 0.69$, and $H_0 = 69.01 \, \mathrm{km\,s^{-1}\,Mpc^{-1}}$.
The corresponding $1\sigma$ confidence intervals, obtained using $\Delta \chi^2 = 2.30$, are
$B_s = 0.687^{+0.0275}_{-0.0303}$ and $H_0 = 69.01^{+2.21}_{-2.20} \, \mathrm{km\,s^{-1}\,Mpc^{-1}}$.

\begin{figure}[h!]
    \centering 
    \begin{tabular}{l l}
        \parbox{2in}{%
            \includegraphics[width=2in,height=2in]{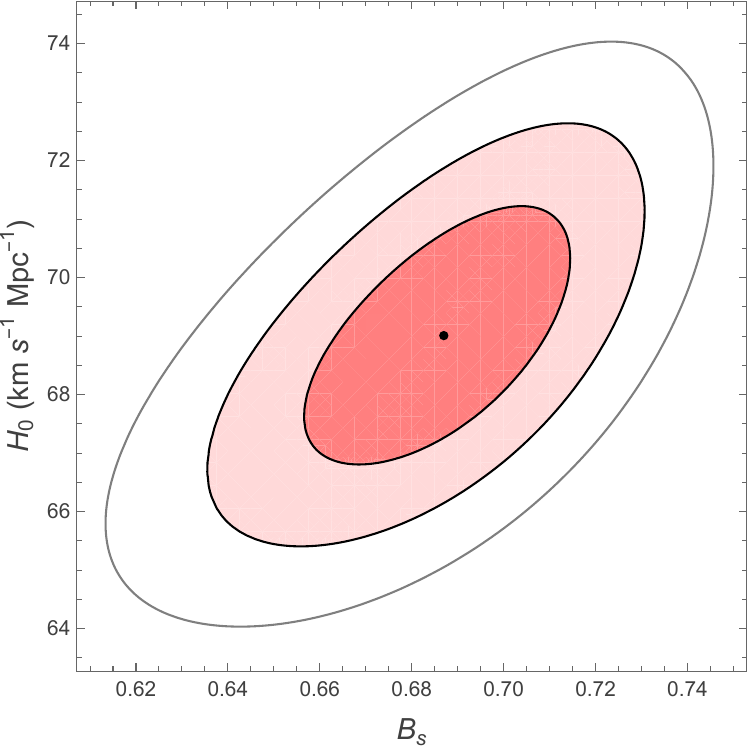}%
        } &
        \parbox{2in}{%
           \includegraphics[width=2in,height=1in]{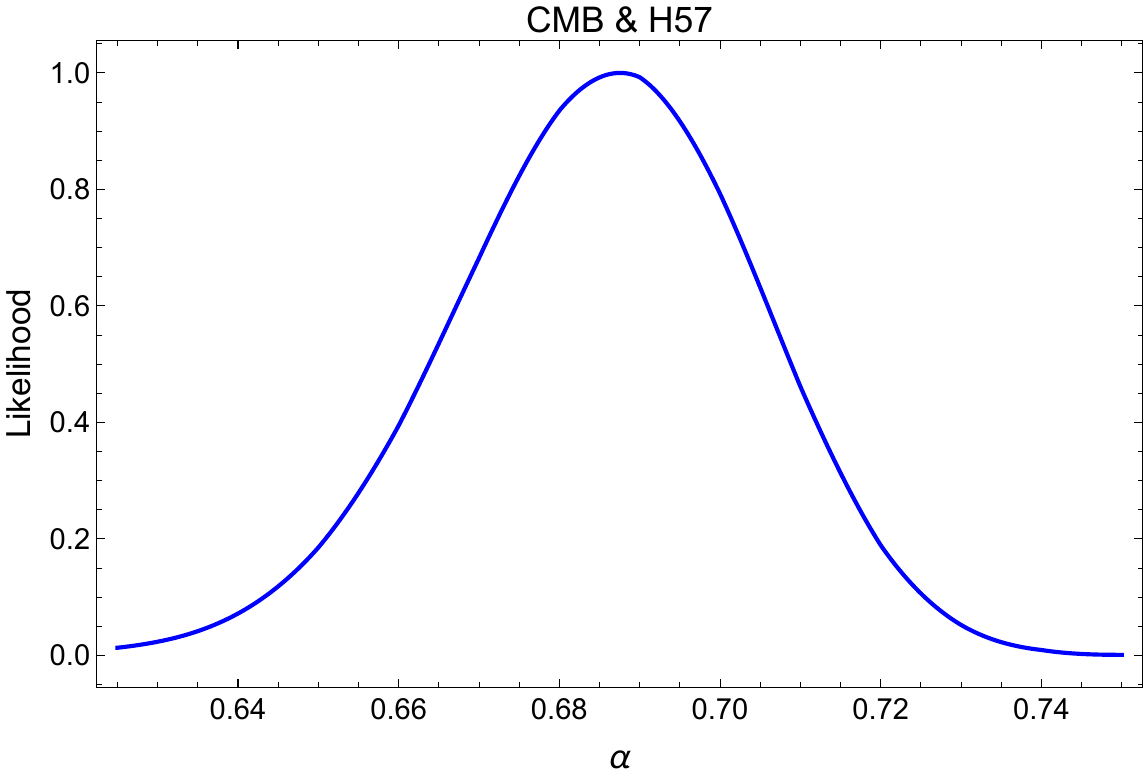}\\
            \includegraphics[width=2in,height=1in]{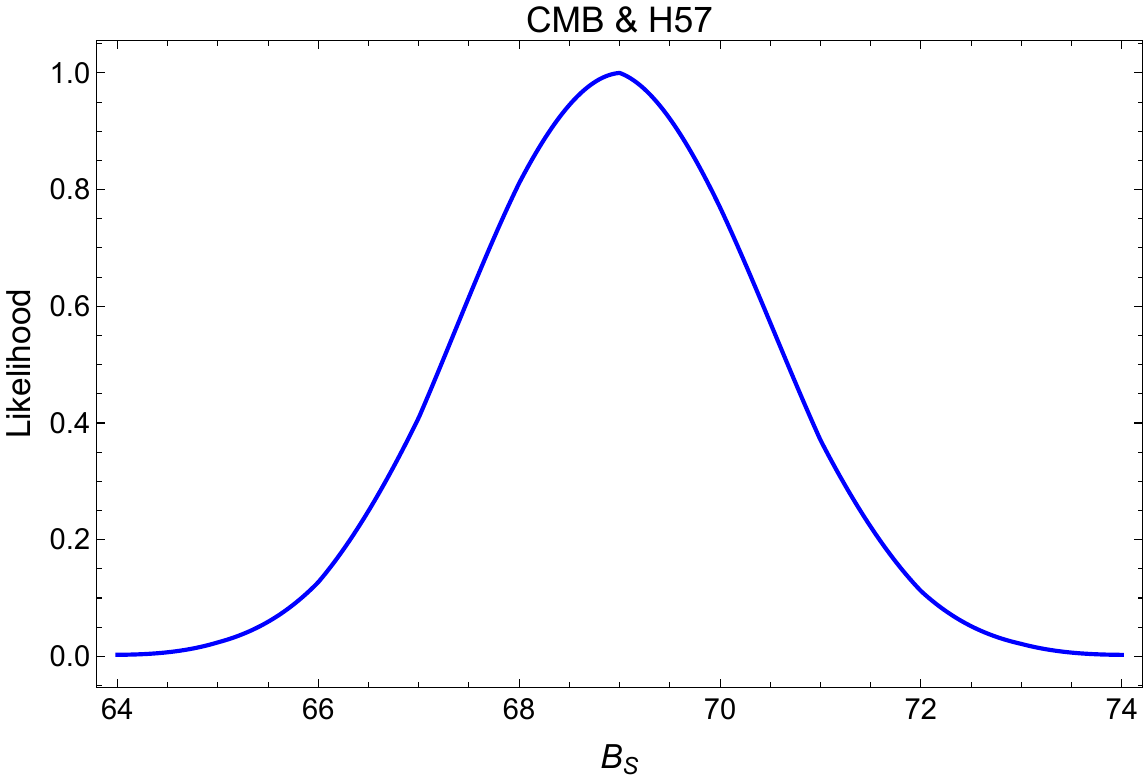}%
        } \\
    \end{tabular}
    \caption{\small\emph{$B_s$ vs $H_0$ graph with liklihood}} 
    \label{mchaa}
\end{figure}

These results show that the inclusion of the CMB constraint substantially improves the parameter estimation by incorporating early-universe information ($z_* \sim 1090$) and reducing degeneracies among the cosmological parameters. The joint contour plots (Fig.~\ref{mchaa}) illustrate the allowed parameter regions at the $1\sigma$, $2\sigma$, and $3\sigma$ confidence levels, demonstrating good agreement between the model and observational data.

It is important to note that, although the first-order approximation is formally applicable only in the late-time regime, the consistency of the model with the combined Hubble-$57$ and CMB analysis over a broad redshift range supports its overall cosmological viability.

\section{Negative Constant ($c < 0$) }
In the previous section, we considered a positive value of the integration constant $c$ to study the evolution of the Modified Chaplygin Gas model. In what follows, we examine the effect of a negative value of $c$ and investigate whether it leads to any significant change in the dynamical evolution of the model.

\subsection{Phantom field}
A distinctly new cosmological behaviour emerges when the arbitrary integration constant is taken to be $c < 0$. Let us define $c = - C$, where $C > 0$. In this case, the energy density increases with the scale factor, thereby mimicking a phantom dark energy scenario. However, at late times it asymptotically approaches a constant value, effectively behaving like a cosmological constant. We get from Eq.~\eqref{eq:7} that for the matter field to be well behaved the
condition
\begin{equation}\label{eq:36}
a^{3(1 +A)(1 + \alpha)}> \frac{C(1 + A)}{B},
\end{equation}
need to be satisfied. So a minimal value of the scale factor given
by
\begin{equation}\label{eq:37}
a(t)_{min}= \left[\frac{C(1 +A)}{B}\right]^{\frac{1}{3(1 +A)(1 + \alpha)}}.
\end{equation}
We consider a homogeneous scalar field $\phi(t)$  and a potential $U(\phi)$ to describe
the Lagrangian as

\begin{equation}\label{eq:38}
L_{\phi} = \frac{\dot{\phi}^2}{2}  - U(\phi),
\end{equation}
and now set the energy density of the field equal to that of the Modified Chaplygin gas as
\begin{equation}\label{eq:39}
\rho_{\phi} = \frac{\dot{\phi}^2}{2}  - U(\phi) =  \rho.
\end{equation}
The corresponding `pressure' coincides with the Lagrangian density as

\begin{equation}\label{eq:40}
p_{\phi} = \frac{\dot{\phi}^2}{2}  - U(\phi) = p.
\end{equation}
These gives
\begin{equation}\label{eq:41}
 \dot{\phi}^2   =  \frac{c(1+A)}{\rho^{\alpha} a^{3(1+A)(1+\alpha)}}.
\end{equation}
When the constant $c < 0$, the kinetic term becomes negative, i.e., $\dot{\phi}^2 < 0$ . In this case, we may define a real scalar field $\psi$ through the relation  $\phi = i \psi$. Accordingly, the Lagrangian of the field $\phi(t)$ can be rewritten as
\begin{equation}\label{eq:41a}
L_{\phi} = \frac{\dot{\phi}^2}{2}  - U(\phi) = - \frac{1}{2} \dot{\psi}^2 - U(i\psi).
\end{equation}
The energy density and the pressure corresponding to the scalar field $\psi$ are as following
\begin{equation}\label{eq:42}
\rho_{\psi} = \frac{\dot{\psi}^2}{2}  + U( i \psi),
\end{equation}
and
\begin{equation}\label{eq:43}
p_{\psi} = \frac{\dot{\psi}^2}{2}  - U( i \psi),
\end{equation}
therefore, the scalar field $\psi$ behaves as a phantom field. This demonstrates that a phantom-like equation of state can be realized from an interacting modified Chaplygin gas dark energy model in a non-flat universe.

The phantom nature of the scalar field also naturally points to the possibility of a nonsingular bouncing cosmology at early times. Such scenarios have been investigated extensively by Setare~\cite{set}. Furthermore, the Chaplygin gas model interpolates between a radiation-dominated phase at small values of the scale factor and a cosmological constant--dominated phase at late times. However, this well-known quartessence picture breaks down when the arbitrary integration constant is negative. Following the approach of Barrow~\cite{bar}, the Chaplygin gas dynamics can be reformulated in terms of a scalar field $\phi$ with a self-interaction potential $U$. For a negative value of the integration constant $c$, the transformation $\phi=i\psi$ is required, and the resulting expressions for the energy density and pressure show that the scalar field acquires a phantom nature.

\subsection{Emergent Universe Scenario in the Phantom Regime}

An interesting feature of the Modified Chaplygin Gas (MCG) model arises when the integration constant takes a negative value, $c<0$, leading to a phantom-like effective equation of state. In this case, Eq.~\eqref{eq:37} indicates the existence of a finite minimum scale factor, implying a nonsingular cosmological evolution. However, the existence of a minimum scale factor alone does not determine the precise nature of the cosmological evolution. To investigate this issue further, we rewrite Eq.~\eqref{eq:8} in the following form:
\begin{equation}\label{eq:44}
\frac{\dot{a}}{a} = \frac{1}{\sqrt{3}}\left[ \frac{B}{1+A} - \frac{C}{a^{3(1+\alpha)(1+A)}}\right]^{\frac{1}{2(1+\alpha)}}, C > 0.
\end{equation}
Equation~\eqref{eq:44} implies the existence of a lower bound on the scale factor, $a_{\min}$, below which the solution is not physically admissible. Furthermore, as $a\rightarrow a_{\min}$, the Hubble parameter tends to zero, whereas for $a\gg a_{\min}$ it approaches a constant value, $3H^2 \simeq \left(\frac{B}{1+A}\right)^{\frac{1}{1+\alpha}}$, corresponding to an asymptotic de Sitter phase. Thus, the model describes a regular cosmological evolution connecting a finite minimum scale factor to a late-time accelerated epoch.

However, these properties alone are insufficient to determine the precise nature of the cosmological evolution. A finite minimum scale factor together with $H\rightarrow0$ may arise in an emergent universe, a bouncing cosmology, or other nonsingular scenarios. Therefore, an explicit determination of the time dependence of the scale factor is required to identify the physical nature of the solution.

Since Eq.~\eqref{eq:44} is highly nonlinear, an exact analytical solution for $a(t)$ is not readily obtainable. To overcome this difficulty, we employ a first-order approximation that yields an explicit expression for the scale factor. As will be shown below, the resulting solution approaches the finite value $a_{\min}$ asymptotically as $t\rightarrow-\infty$, while $\dot a\rightarrow0$, indicating a past-eternal quasi-static state. The universe subsequently evolves smoothly into an expanding phase without encountering a Big-Bang singularity. Therefore, the explicit time evolution of the scale factor identifies the present solution as an emergent universe rather than a bouncing cosmology.

Since, in an expanding universe, the scale factor increases with time, we can apply a first-order binomial approximation to Eq.~\eqref{eq:44}, which can be expressed as
\begin{equation}\label{eq:45}
\frac{\dot{a}}{a}= \frac{1}{\sqrt{3}} \left(\frac{B}{1+A}\right)^{\frac{1}{2(1+\alpha)}} - \frac{1}{2 \sqrt{3}(1+\alpha)} \left(\frac{1+A}{B}\right)^{\frac{1 +2\alpha}{2(1+\alpha)}}~\frac{C}{a^{3(1+A)(1+\alpha)}}.
\end{equation}
The general solution of the Eq.~\eqref{eq:45} is obtained as
\begin{equation}\label{eq:46}
a(t) = \left[
\frac{Cd + 2\sqrt{3}(1+ \alpha)\left(\frac{B}{1+A}\right)^{\frac{1+ 2 \alpha}{2(1+\alpha)}} d \, e^{\sqrt{3}(1+\alpha)(1+A)\left(\frac{B}{1+A}\right)^{\frac{1}{2(1+\alpha)}}t}}
{2\left(\frac{B}{1+A}\right)(1+\alpha)}
\right]^{\frac{1}{3(1+\alpha)(1+A)}},
\end{equation}
where $d$ is an integration constant.
\begin{figure}[ht]
\begin{center}
    \includegraphics[width=7.2 cm]{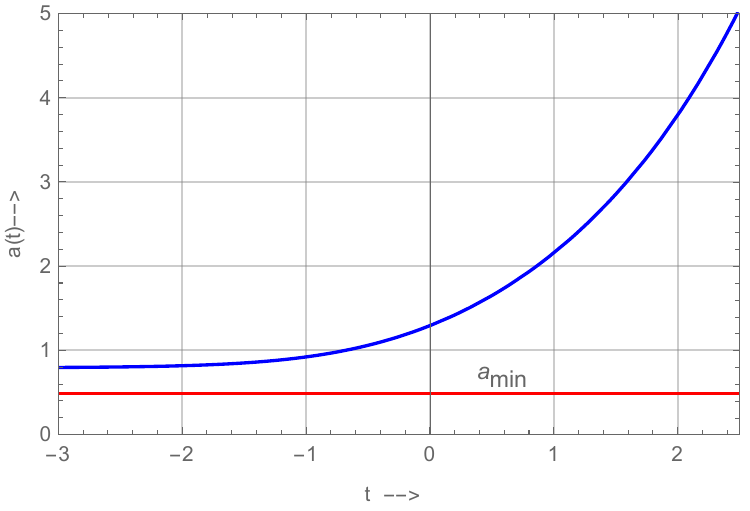}
      \caption{
  \small\emph{The variations of $ a(t)$ and $t$. }\label{ae}
    }
\end{center}
\end{figure}
The above expression resembles the well-known emergent form
\begin{equation}\label{eq:47}
a(t) = \left(\beta + \gamma e^{m t} \right)^{N},
\end{equation}
where
$\beta = \frac{C(1+A)d}{2B(1+\alpha)}$, \qquad
$\gamma = \sqrt{3}d \left(\frac{B}{1+A}\right)^{-\frac{1}{2(1+\alpha)}}$, \qquad
$m = \sqrt{3}(1+\alpha)(1+A)\left(\frac{B}{1+A} \right)^{\frac{1}{2(1+\alpha)}}$, \qquad \\
and $N = \frac{1}{3(1+\alpha)(1+A)}$.

Eq.~\eqref{eq:46} clearly demonstrates that an emergent–universe–type solution can naturally arise within the Modified Chaplygin Gas (MCG) framework under the first-order approximation (Fig.~\ref{ae}). It is worth noting that S. Mukherjee \textit{et al.}~\cite{sm} previously derived an emergent universe solution by adopting $\alpha = -\tfrac{1}{2}$ in Eq.~\eqref{eq:1c}. In contrast, S. Dutta \textit{et~al.}~\cite{sd} reported that an emergent-type solution does not arise from the conventional MCG model. The present solution, however, described by Eq.~\eqref{eq:46}, represents a new class of emergent-type behavior that naturally originates from the MCG equation of state under the first-order approximation. Interestingly, for $c < 0$, the model corresponds to a phantom regime, provided the physically admissible conditions $0 < A < \tfrac{1}{3}$ and $0 < \alpha < 1$ are satisfied. We now  discuss the cosmological implications of this solution.

When $c < 0$, Eq.~\eqref{eq:44} imposes a lower bound on the scale factor to maintain a positive energy density. This condition leads directly to emergent behaviour. To ensure the positivity of the density for negative $c$
the scale factor cannot shrink below a finite minimum value, $a(t)_{min}$, as given in Eq.~\eqref{eq:37} which guarantees that the universe never reaches a singular state ($a=0$).

Taking the limit $t \to -\infty$ in Eq.~\eqref{eq:46}, the exponential term vanishes and the scale factor tends to a constant:
\begin{equation}\label{eq:48}
\lim_{t \to -\infty} a(t) = \left[ \frac{C(1+A)d}{2B(1+\alpha)} \right]^{\frac{1}{3(1+\alpha)(1+A)}},
\end{equation}
which remains consistent with Eq.~\eqref{eq:37} for $d = 2(1 + \alpha)$. Hence, as $t \to -\infty$, the universe approaches a static configuration with a finite minimum radius $a_{\min}$. Subsequently, it enters an exponentially expanding phase, exhibiting a de Sitter–like behavior. With further cosmic evolution, as $t \to +\infty$, the expansion continues to accelerate, and in the asymptotic limit, Eq.~\eqref{eq:46} reduces to $a(t) \sim e^{Ht}$, corresponding to the exponential expansion characteristic of the $\Lambda$CDM model. Thus, the model presents a smooth and singularity-free cosmological evolution, beginning from a quasi-static past and naturally evolving into a late-time accelerating universe consistent with $\Lambda$CDM dynamics.

For $c<0$, this correspondence shows that the emergent behaviour in the MCG framework is dynamically equivalent to that of a universe dominated by a phantom scalar field, driving cosmic acceleration without an initial singularity in first approximation. The emergent universe described above is non-singular, beginning from a quasi-static state with finite $a(t)_{\min}$ and gradually evolving into an accelerating phase dominated by dark-energy-like behaviour. The transition is smooth, and the absence of an initial singularity makes this model theoretically attractive. Therefore, within the MCG scenario, the case $c<0$ (phantom regime) provides a consistent and physically meaningful realization of an emergent universe.

The energy density of the universe is given by
\begin{equation}\label{eq:49}
\rho = \left(\frac{B}{1+A}\right)^{\frac{1}{1+\alpha}}\left(\frac{\gamma e^{mt}}{\beta + \gamma e^{mt}}\right)^2,
\end{equation}
which remains positive during the cosmic evolution. At early times ($t \to -\infty$), $e^{m t} \to 0$, and hence $\rho \to 0$, indicating a quasi-static phase with negligible energy density. As time progresses, $e^{m t}$ increases, leading to a monotonic growth of $\rho(t)$.
In the late-time limit ($t \to +\infty$), $e^{m t} \gg \beta$, and the energy density approaches a constant value
$\rho \to \left(\frac{B}{1+A} \right)^{\frac{1}{1+\alpha}} $, representing a dark energy–dominated de Sitter phase. Thus, the evolution of $\rho(t)$ shows a smooth transition from a low-energy emergent state in the past to a constant energy density at late times, consistent with the observed accelerated expansion of the universe.
The corresponding deceleration parameter $q$ is expressed as
\begin{equation}\label{eq:50}
q = -1 - \frac{\beta}{N \gamma} e^{-m t},
\end{equation}
which remains negative throughout the cosmic evolution, indicating that the universe is always in an accelerated phase. This time-varying deceleration parameter~\cite{vs} signifies that, although the universe continues to accelerate, the strength of this acceleration gradually decreases with cosmic time.
Furthermore, the time derivative of $q$ is
\begin{equation}\label{eq:50a}
\frac{dq}{dt} = m \frac{\beta}{N \gamma} e^{-m t},
\end{equation}
which is positive for $\beta, m, N, \gamma > 0$, confirming that $q(t)$ steadily increases with time and, as a result, the acceleration weakens as the universe evolves. Consequently, $q(t)$ evolves from $-\infty$ at early times ($t \to -\infty$) to its asymptotic value $q \to -1$ at late times ($t \to \infty$). This behaviour characterizes a smooth transition toward a de~Sitter--like accelerated phase in the far future.

The effective equation of state is given by
\begin{equation}\label{eq:51}
w_{\mathrm{eff}}(t) = A - (1+A)\left(1+\frac{\beta}{\gamma}e^{-mt}\right)^{2(1+\alpha)}.
\end{equation}
As $t \to -\infty$, $w_{\mathrm{eff}}(t) \to -\infty$, since the pressure term $-\frac{B}{\rho^{\alpha}}$  of Eq.~\eqref{eq:1c} dominates while $\rho \to 0$, and the universe remains quasi-static.

As $t \to \infty$, $w_{\mathrm{eff}}(t) \approx -1$. Therefore, $w_{\mathrm{eff}}(t)$ monotonically increases from $-\infty$ toward its late-time asymptotic value.   Physically, this behavior of $w_{\mathrm{eff}}(t)$ captures the evolution of an emergent universe scenario. At early times ($t \to -\infty$), $w_{\mathrm{eff}} \to -\infty$ reflects the dominance of the negative pressure term, keeping the universe in a quasi-static state and avoiding an initial singularity. As time progresses, $w_{\mathrm{eff}}(t)$ increases monotonically, approaching a value close to $-1$ at late times, which drives the accelerated expansion observed in the present epoch. This smooth evolution from a quasi-static past to a late-time accelerated expansion ensures a singularity-free, continuous cosmic history, consistent with the emergent universe picture.

Equation~\eqref{eq:46} demonstrates that an emergent–universe–type solution naturally arises within the Modified Chaplygin Gas (MCG) framework under the first-order approximation. While S.~Mukherjee \textit{et~al.}~\cite{sm} obtained such a solution for $\alpha = -\tfrac{1}{2}$, S.~Dutta \textit{et~al.}~\cite{sd} reported its absence in the conventional MCG model. In contrast, the present solution corresponds to a new class of emergent-type behavior arising for $c < 0$ (phantom regime) within the physically admissible range $0 < A < \tfrac{1}{3}$ and $0 < \alpha < 1$.

At the present epoch ($a_0 = 1$), the energy density and deceleration parameter are given by
\begin{equation}\label{eq:51a}
\rho_0 = \left(\frac{B}{1+A}\right)^{\frac{1}{1+\alpha}}(1 - \beta)^2, \qquad
q_0 = -1 - \frac{\beta}{N (1-\beta)},
\end{equation}
which clearly indicates an accelerating universe. To ensure this behavior, the condition $\beta < 1$ must be satisfied, which further implies that $B > \dfrac{C(1+A)d}{2(1+\alpha)}$.
The effective equation of state at the present epoch is then given by
$w_{\mathrm{eff},0} = A - \frac{1+A}{(1-\beta)^{2(1+\alpha)}}$, which becomes negative for $0<\beta < 1$ and $0< \alpha < 1$. This result signifies that the present universe is dominated by dark energy, in agreement with observational evidence for the ongoing accelerated expansion.

It is worth noting that the present emergent universe solution corresponds to the case $c<0$, for which the scale factor approaches a finite minimum value $a_{\min}$ as $t\rightarrow -\infty$. Consequently, the universe never reaches the singular state $a=0$. If a radiation component satisfying $\rho_r\propto a^{-4}$ is included, the radiation density also remains finite since $a\ge a_{\min}>0$. Therefore, the inclusion of radiation does not reintroduce an initial singularity or alter the emergent nature of the solution.

However, a realistic description of primordial processes such as Big Bang nucleosynthesis generally requires the presence of a radiation-dominated epoch. This may be achieved by extending the present framework to include an explicit radiative fluid together with the Modified Chaplygin Gas component. In such a scenario, radiation can naturally dominate the energy density during the early stages of expansion and recover the standard thermal history required for primordial nucleosynthesis, while the MCG component provides the nonsingular emergent background.

A detailed investigation of nucleosynthesis constraints has not been carried out in the present work, since such an analysis requires not only the inclusion of radiation but also a careful treatment of the thermal history and the nuclear reaction network governing the production of light elements. Since the primary objective of this work is to study the background dynamics of the Modified Chaplygin Gas model, its emergent universe solution, and the associated observational constraints, a dedicated nucleosynthesis analysis lies beyond the scope of the present work. Nevertheless, the present results indicate that the emergent MCG scenario may consistently accommodate an additional radiative fluid, which is required for a realistic description of the nucleosynthesis epoch while preserving the nonsingular character of the cosmological evolution.

In the next subsection we shall examine whether the cosmological solution remains regular throughout its evolution and the completeness of geodesic.

\subsection{Geodesic Completeness and Regularity of the Emergent Universe}

\noindent
A key requirement for any viable emergent universe model is the absence of spacetime singularities and the completeness of geodesics. In this context, it is essential to examine whether the cosmological solution remains regular throughout its evolution and whether all timelike and null geodesics can be extended to arbitrary values of the affine parameter. This provides a robust criterion for distinguishing a genuinely non-singular emergent scenario from models that merely avoid an initial singularity at the level of the scale factor.

\medskip

To assess the physical viability of the emergent universe solution corresponding to the scale factor~\eqref{eq:47} within the spatially flat FLRW metric~\eqref{eq:1}, we analyze both the geodesic structure and the behavior of curvature invariants. We know the motion of particles is governed by the geodesic equation
\begin{equation}\label{eq:51b}
\frac{d^2 x^\mu}{d\lambda^2} + \Gamma^\mu_{\nu\rho} \frac{dx^\nu}{d\lambda}\frac{dx^\rho}{d\lambda} = 0,
\end{equation}
where $\lambda$ is the affine parameter. A spacetime is geodesically complete if the affine parameter $\lambda$  can be extended to $t \pm \infty$
along all geodesics (timelike and null).

\medskip

\textit{(i) Null geodesics:}

For radial null geodesics ($ds^2=0$), we obtain
\begin{equation}\label{eq:51c}
\frac{dr}{dt} = \frac{1}{a(t)}.
\end{equation}
Using Eqs ~\eqref{eq:51b} \& ~\eqref{eq:51c} we obtain
\begin{equation}\label{eq:51d}
\frac{dt}{d\lambda} \propto \frac{1}{a(t)}.
\end{equation}
Thus, the affine parameter behaves as $ \lambda \propto \int a(t)\, dt $.
Therefore, the completeness of null geodesics depends on the behavior of the integral
\begin{equation}\label{eq:51e}
\lambda \propto \int (\beta+\gamma e^{mt})^N dt.
\end{equation}

\noindent
In the asymptotic past ($t \to -\infty$), we have $e^{mt} \to 0$, and hence $ a(t) \to \beta^N = a_{\min} \neq 0 $, which leads to $ \lambda \propto t$, indicating that the affine parameter diverges. Therefore, null geodesics are complete in the past. In the late-time limit ($t \to \infty$), the scale factor behaves as $ a(t) \sim \gamma^N e^{mN t} \sim  \gamma^N e^{H t} $. Consequently, the relevant integral behaves as $ \int a(t) dt \sim \int e^{mN t} dt $, which also diverges. This reflects the asymptotically de Sitter nature of the spacetime. A null geodesic is considered regular if $\lambda \rightarrow \pm\infty $ as $t \rightarrow \pm \infty$. However, since no curvature divergence occurs and the spacetime remains regular, this behavior does not indicate a physical singularity. The divergence of the affine parameter in the asymptotic past, together with the regularity of the spacetime at late times, implies that the geodesics can be extended indefinitely. Combined with the finiteness of curvature invariants, this confirms that the spacetime is geodesically complete and free from physical singularities.
\medskip

\textit{(ii) Timelike geodesics:}

For comoving observers, the proper time satisfies
\begin{equation}\label{eq:51f}
d\tau = dt \quad \Rightarrow \quad \tau = \int dt,
\end{equation}
which diverges as $t \to \pm \infty$. Thus, timelike geodesics are also complete.
\medskip

\noindent
Now the Hubble parameter is given by
\begin{equation}\label{eq:51g}
H = \frac{\dot a}{a} = \frac{mN\gamma e^{mt}}{\beta+\gamma e^{mt}},
\end{equation}
which remains finite for all $t$ and approaches a constant value $H \to mN $ where $ mN  = \frac{1}{\sqrt{3}} \left(\frac {B}{1+A}\right)^{\frac{1}{2(1+\alpha)}} $ at late times.  The Ricci scalar is given by for metric~\eqref{eq:1},
\begin{equation}\label{eq:51h}
R = 6\left(\dot H + 2H^2\right) = \frac{6 m^2 \gamma N e^{mt}}{(\beta + \gamma e^{mt})^2}
\left[  \beta  + 2N \gamma e^{mt}\right].
\end{equation}

\noindent
Now the Kretschmann scalar is given by
\begin{equation}\label{eq:51i}
K = R_{\mu\nu\rho\sigma}R^{\mu\nu\rho\sigma}
= 12\left\{H^4 + (\dot H + H^2)^2\right\}  = \frac{12 m^4 N^2 \gamma^2 e^{2mt}}{(\beta + \gamma e^{mt})^4}
\left[ N^2 \gamma^2 e^{2mt} + \left( \beta   + N \gamma e^{mt}\right)^2 \right].
\end{equation}

\noindent
In the asymptotic past ($t \to -\infty$), we find $ R \to 0 $ and $  K \to 0 $, indicating a regular, non-singular initial state. In the late-time limit ($t \to \infty$), the spacetime approaches a de Sitter phase with $ R \to 12 (mN)^2 $ and $ K \to 24 (mN)^4 $ which are finite constants.

So the model exhibits a finite minimum scale factor in the infinite past, asymptotically approaches a de Sitter phase in the future, and maintains finite curvature invariants throughout its evolution. Timelike geodesics are complete, and null geodesics do not encounter any curvature singularity. Therefore, the spacetime represents a consistent non-singular emergent universe scenario.

In the next subsection, we carry out an observational constraint analysis using the Hubble-$57$ dataset in combination with the CMB shift parameter, in order to determine the allowed ranges of the model parameters and to examine the viability of the emergent universe scenario within the modified Chaplygin gas (MCG) framework.

\subsection{Observational Constraints}

Here, as in the previous cases, we constrain the model parameters using the Hubble-$57$ and combined Hubble-$57$ \& CMB analyses.

\subsubsection{Constraints from Hubble-$57$ dataset}
\begin{figure}[h!]
    \centering 
    \begin{tabular}{l l}
        \parbox{2in}{%
            \includegraphics[width=2in,height=2 in]{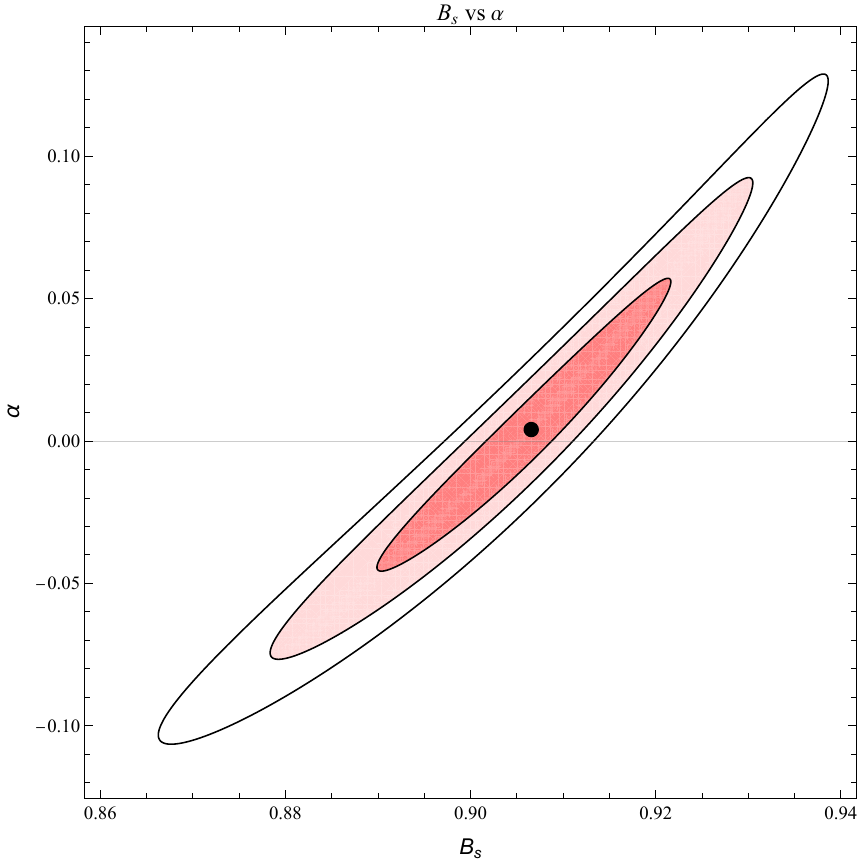}%

        } &
        \parbox{2in}{%
           \includegraphics[width=2in,height=1in]{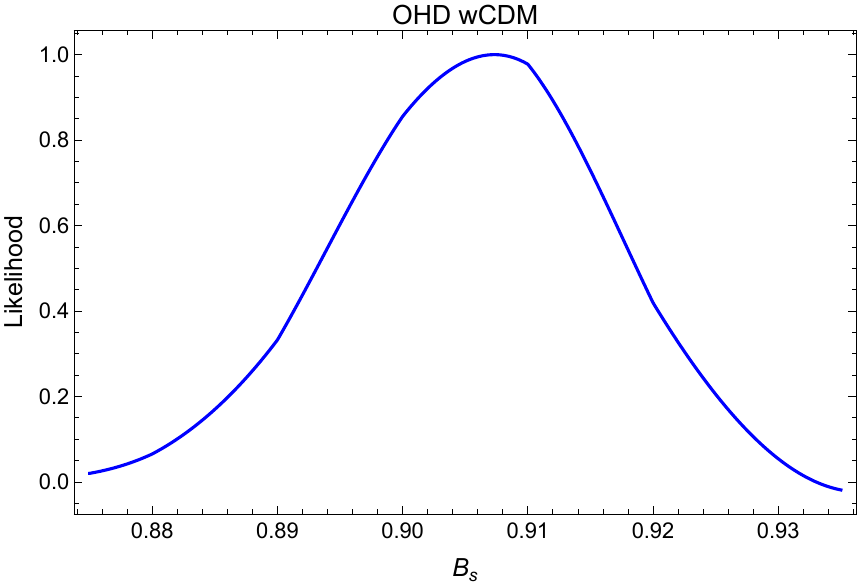}\\
            \includegraphics[width=2in,height=1in]{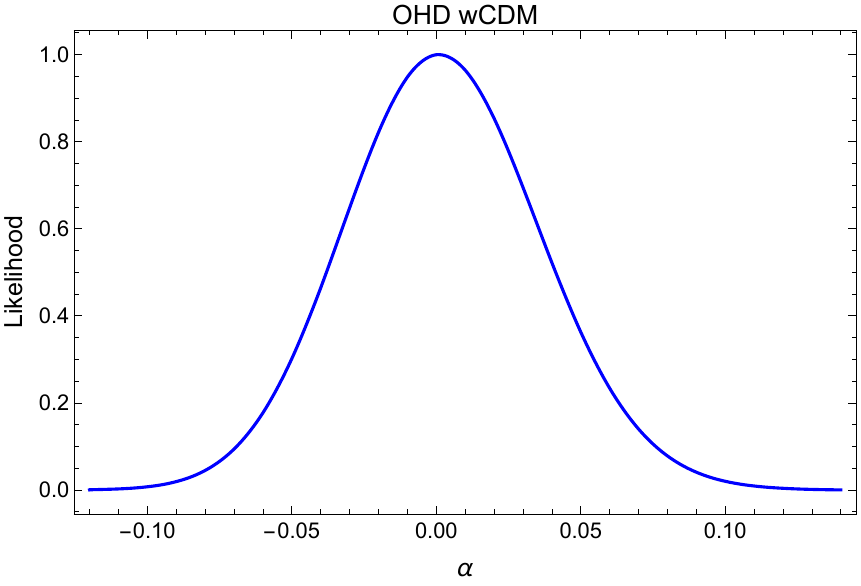}%
        } \\
    \end{tabular}
    \caption{\small\emph{$B_s$ vs $\alpha$ graph with liklihood}} 
    \label{Lbsc}
\end{figure}

To constrain the model parameters, we previously considered
 $B_s = \left({\frac{B}{1+A}}\right)^{\frac{1}{1+\alpha}}\frac{1}{\rho_0}$ and $C = -(1-B_s^{1+\alpha})\rho_0^{1+\alpha}$, we get the expression of Hubble parameter in terms of $B_s$ from Eq.~\eqref{eq:45} as

\begin{equation}\label{eq:52}
H = H_0 B_s^{\frac{1}{2}} \left \{ 1  +\frac{1}{2(1+\alpha)}\frac{\left(1- B_s^{1+\alpha} \right)}{B_s^{1+\alpha}}\left(1+z \right)^{3(1+A)(1+\alpha)}
\right\}.
\end{equation}
The values of $B_s$ and $\alpha$ corresponding to $\chi^2_{m}$ are determined from the contour plot (see Fig.~\ref{Lbsc}) and  their respective ranges are given in Table~\ref{t5}, for $H_0 = 80.2~kms^{-1} Mpc^{-1}$ and $A = 0.003$.
 We observe that $0<\alpha < 1$, in contrast to the earlier work of S. Mukherjee \textit{et al.}~\cite{sm} on the emergent universe model, where the parameter was taken as $\alpha = -\tfrac{1}{2}$.

\begin{table}[h!]
\centering
\caption{Best-fit values of the $B_s$ and $\alpha$ using Hubble-$57$ data (1$\sigma$ confidence region).}
\resizebox{\textwidth}{!}{%
\begin{tabular}{lccccc}
\hline
\textbf{Parameter} & $H_0~(\mathrm{km\,s^{-1}\,Mpc^{-1}})$ & $A$ & $\chi^2_m$ & $B_s$ & $\alpha$ \\
\hline
\textbf{Best-fit values} & 80.20 & 0.003 & 84.49 & 0.91 & 0.0044 \\
\textbf{Range (1$\sigma$ region)} & -- & -- & -- & $0.8899$, $0.9216$ & $-0.0457$, $0.0571$ \\
\hline
\end{tabular}
}
\label{t5}
\end{table}
It is worth emphasizing that the condition $\beta < 1$, derived in Sec.~5.2, upon substituting $d = 2(1+\alpha)$, reduces to the inequality $\frac{C(1+A)}{B} < 1$. This relation, in turn, implies $\frac{1}{B_s^{1+\alpha}} > 0$, and hence $B_s$ should be positive. From Table~5, we obtain $B_s = 0.91$, which clearly satisfies this requirement and remains fully consistent with the parameter constraints obtained from Eq.~\eqref{eq:52} at $\chi_m^2$.\\

We also determine the present age of the universe using the parameter values from Table~\ref{t5}. From Eqs.~\eqref{eq:21} and \eqref{eq:52}, we obtain the present age as $t_0 = 12.92$~Gyr, which is slightly lower than the value reported by the Planck 2020 results. This comparatively lower age arises primarily due to the higher Hubble parameter ($H_0 = 80.2~\mathrm{km,s^{-1},Mpc^{-1}}$) and the continuous accelerated expansion associated with the $c < 0$ emergent–universe scenario, both contributing to a faster overall expansion rate of the universe.

\subsubsection{Constraints from CMB Shift Parameter and Hubble-$57$ Data}
\noindent Now using Eq.~\eqref{eq:52}, for the adopted parameter values $A=0.003$, $\alpha=0.00135$, $\Omega_m = 0.30$ and $B_s=0.952$, we obtain $R_{\mathrm{th}} = 1.741$, which is in good agreement with the observed value. The corresponding chi-square value, $\chi^2_{\mathrm{CMB}} = 1.04$, indicates consistency within the $1\sigma$ confidence level, thereby supporting the compatibility of the model with early-universe observations.

\noindent To obtain improved constraints, we perform a combined Hubble-$57$ \& CMB analysis by fixing $A = 0.003$, $\alpha = 0.00135$, and $\Omega_m = 0.30$, while treating $B_s$ and $H_0$ as free parameters. Minimization of the total chi-square function, $\chi^2_{\mathrm{total}} = \chi^2_H  + chi^2_{\mathrm{CMB}}$, yields $\chi^2_{\mathrm{min}} = 65.69$, with best-fit values $B_s = 0.952$ and $H_0 = 83.25 \, \mathrm{km\,s^{-1}\,Mpc^{-1}}$. The corresponding $1\sigma$ confidence intervals, obtained using $\Delta \chi^2 = 2.30$, are $B_s = 0.952^{+0.0010}_{-0.0011}$ and $H_0 = 83.25^{+2.94}_{-2.94} \, \mathrm{km\,s^{-1}\,Mpc^{-1}}$.

\begin{figure}[h!]
    \centering 
    \begin{tabular}{l l}
        \parbox{2in}{%
            \includegraphics[width=2in,height=2in]{cc}%

        } &
        \parbox{2in}{%
           \includegraphics[width=2in,height=1in]{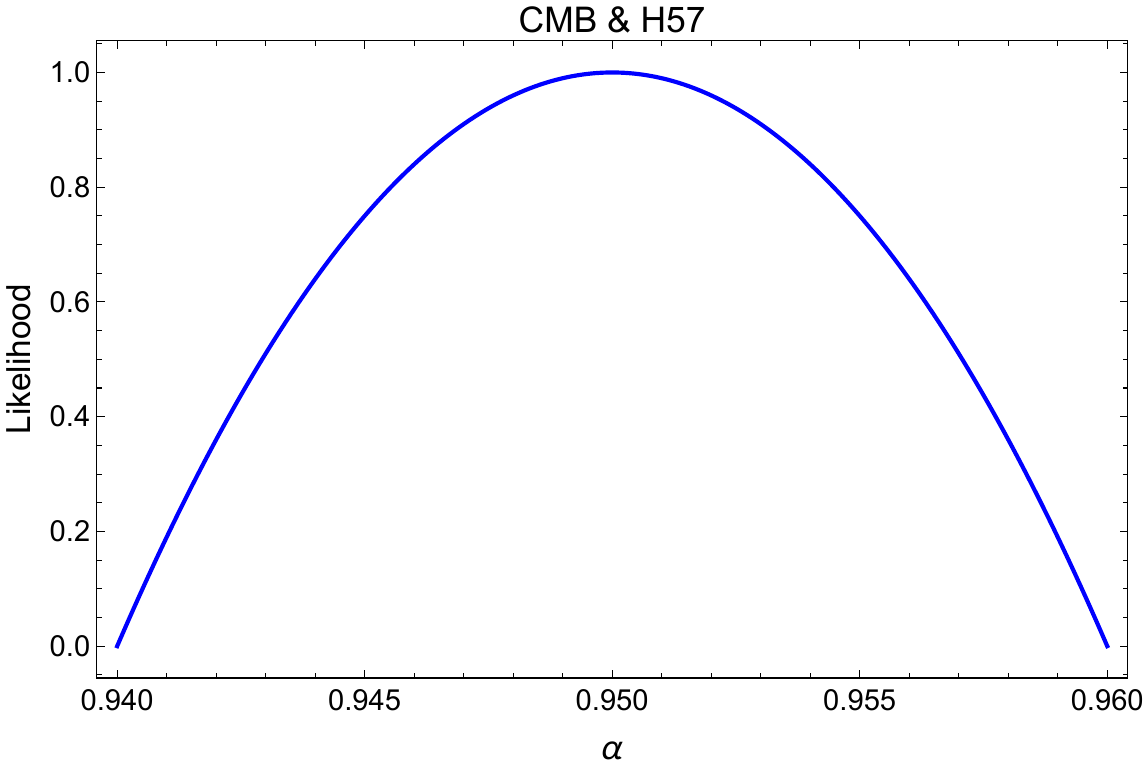}\\
            \includegraphics[width=2in,height=1in]{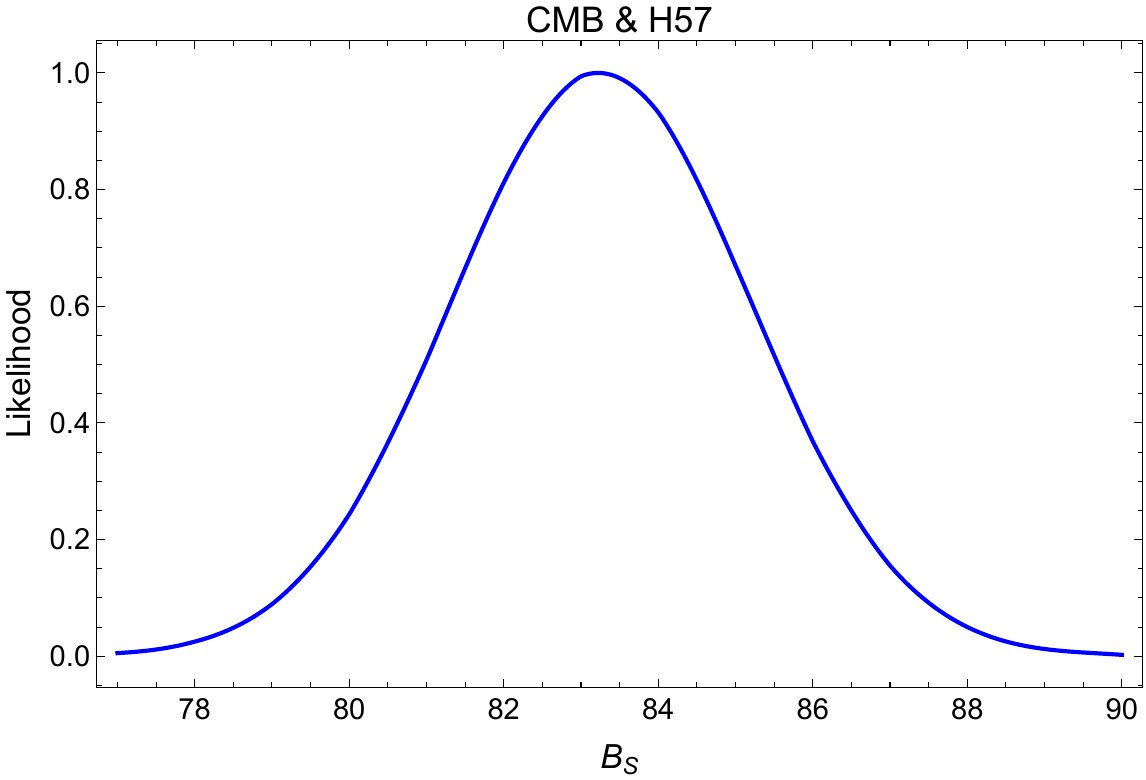}%
        } \\
    \end{tabular}
    \caption{\small\emph{$B_s$ vs $H_0$ graph with liklihood}} 
    \label{ch}
\end{figure}

\noindent The best-fit value of $B_s$ is close to unity, indicating that the model effectively mimics a dark energy dominated universe at late times, with tight constraints from the combined Hubble-$57$ and CMB data. The estimated Hubble parameter, $H_0 \approx 83.25 \, \mathrm{km\,s^{-1}\,Mpc^{-1}}$, is higher than the \textit{Planck} value but closer to local measurements, with a moderate uncertainty reflecting sensitivity to late-time expansion. The inclusion of the CMB shift parameter significantly improves the constraints by incorporating early-universe information ($z_* \sim 1090$), thereby reducing parameter degeneracies. The joint contour plots (Fig.~\ref{ch}) illustrate the allowed parameter space through the $1\sigma$, $2\sigma$, and $3\sigma$ confidence regions.

It is important to note that the first-order approximation adopted in this work is valid only in the large scale factor regime and therefore primarily describes the late-time evolution of the model. Within this framework, the emergent universe behavior obtained for $c<0$ should be interpreted as a late-time feature rather than a complete description of the primordial universe. However, the combined Hubble-$57$ and CMB analysis, which incorporates both low- and high-redshift observations ($z_* \sim 1090$), demonstrates that the model remains consistent across the entire cosmic evolution. This ensures that the overall viability of the model is not limited by the domain of validity of the approximation.

Overall, the combined analysis shows that the model provides a coherent and self-consistent description of cosmic evolution, successfully accommodating both early- and late-time observational constraints.

\section{Raychaudhuri Equation}

It may not be out of place to address and compare the situation
discussed in the last section with the help of the well known Raychaudhuri equation~ \cite{ray},  which in general holds for any
cosmological solution based on Einstein's gravitational field
equations. With matter field expressed in terms of mass density
and pressure Raychaudhury equation  reduces to a compact form as
\begin{equation}\label{eq:53}
  \dot{\theta}=-2(\sigma^{2}-\omega^{2})-\frac{1}{3}\theta^{2}-\frac{8\pi G}{2}\left(\rho+3p \right),
\end{equation}
in a  co moving reference frame. Here $p$ is the
 isotropic pressure and $\rho$ is the energy density from varied sources.
\vspace{0.1 cm} Moreover other quantities are defined with the
help of a unit vector $v^{\mu}$ as under

\begin{subequations}\label{eq:54}
\begin{align}
\mathrm{the ~expansion ~scalar}~~\theta & =  v^{i}; _{i}  \label{second}\\
 \sigma^{2} & =  \sigma_{ij}\sigma^{ij}  \label{third}\\
\mathrm{the~ shear ~tensor} ~~~~~~\sigma_{ij} &= \frac{1}{2}(v_{i;j}+ v_{j;i})- \frac{1}{2}(\dot{v}_{i}v_{j}+\dot{v}_{j}v_{i})-
\frac{1}{3}v^{\alpha}_{;\alpha}(g_{ij} - v_{i}v_{j})  \label{fourth} \\
\mathrm{the~ vorticity~ tensor}~  \omega_{ij} & =
\frac{1}{2}(v_{i;j}- v_{j;i})
-\frac{1}{2}(\dot{v}_{i}v_{j}-\dot{v}_{j}v_{i}) \label{fifth}
\end{align}
\end{subequations}
  We can calculate  an expression for effective
 deceleration parameter as
\begin{equation}\label{eq:55}
 q= -\frac{\dot{H} + H^{2}}{H^{2}} = -1-3~\frac{\dot{\theta}}{\theta^{2}}.
\end{equation}
 In our case as we are dealing with an isotropic rotation free
spacetime both  the shear  and vorticity scalars vanish which allows us to write,
\begin{equation}\label{eq:56}
  \theta^{2}q =  12 \pi G \left( \rho   + 3p \right).
\end{equation}

Case 1: With the help of the Eqs~\eqref{eq:1c}, \eqref{eq:7} and \eqref{eq:56}  we finally get,

\begin{equation}\label{eq:57}
  \theta^{2}q =  12 \pi G \left[  (1+3A) - \frac{3(1+A)(1-\Omega_m)}{(1-\Omega_m) + \Omega_m(1+z)^{3(1+\alpha)(1+ A)}}   \right].
 \end{equation}
It follows from the Eq.~\eqref{eq:57} that flip occurs (i.e., at $q=0$) when

\begin{equation}\label{eq:58}
z_f = \left\{\frac{2(1-\Omega_m)}{\Omega_m(1+3A)}\right\}^{\frac{1}{3(1+\alpha)(1+A)}} -1.
\end{equation}
This is identical to Eq.~\eqref{eq:13}. Moreover, setting $A = 0$ reduces the Modified Chaplygin Gas (MCG) model to the Generalized Chaplygin Gas (GCG) model, yielding an expression identical to that obtained in our previous work~\cite{dp25}.\\

\textbf{Case 2:} Now we have discussed our alternative approach in the context of Raychaudhuri equation. Now using Eqs~\eqref{eq:24}, \eqref{eq:25} and \eqref{eq:56}  we finally get after straight forward calculation that

\begin{equation}\label{eq:59}
\theta^{2}q =  72 \pi G n \omega^2 \frac{ (1 - n \cosh^2\omega t)}{\sinh^2 \omega t} = 72 \pi G n \omega^2\left[ \frac{1}{n\{1+(1+z)^{-\frac{2}{n}}\}}-1\right].
\end{equation}

We may calculate the flip time from Eq.~\eqref{eq:59} as

\begin{equation}\label{eq:60}
t_{f} = \frac{1}{\omega} \cosh^{-1} \left(\sqrt{\frac{1}{n}} \right) = \frac{1}{\omega} cosh^{-1}\sqrt{\frac{3(1+\alpha)(1+A)}{2}},
\end{equation}
which is identical with the  Eq.~\eqref{eq:27}.

Also, from Eq.~\eqref{eq:59}, we can derive the expression for the redshift at flip time $z_f$,
which is given by

\begin{equation}\label{eq:61}
z_f = \left(\frac{n}{1-n}\right)^{\frac{n}{2}} - 1.
\end{equation}

The derived expression for the redshift at the flip time $z_f$ in Eq.~\eqref{eq:61}  matches  perfectly with earlier findings, as shown in Eq.~\eqref{eq:27a}.

\textbf{Case 3:} Using Eqs~\eqref{eq:1c} and \eqref{eq:56} we get

\begin{equation}\label{eq:62}
\theta^2 q = \frac{3}{2}\rho \left[ 1 + 3\left(A- \frac{B}{\rho^{\alpha+1}}\right)\right].
\end{equation}
Since, $\theta = 3H$ and $\beta << 1$, at present epoch ($a_0 =1$) we finally get

\begin{equation}\label{eq:63}
q_0 \approx -1 - \frac{1}{N} \frac{\beta}{1-\beta},
\end{equation}
which is identical with Eq.~\eqref{eq:51a}.

The consistent results obtained from all approaches demonstrate the robustness of the methodology
and lend further credibility to the theoretical framework adopted in this analysis.

\section{Concluding Remarks}

We explored the emergent universe scenario in general relativity with the modified Chaplygin gas (MCG) for the equation-of-state parameters $0 < A < \tfrac{1}{3}$ and $0 < \alpha < 1$. The resulting cosmological model admits late-time cosmic acceleration within the framework of a spherically symmetric homogeneous universe, with MCG serving as the dark energy component. The field equation, Eq.~\eqref{eq:8}, is highly nonlinear and cannot be solved exactly in closed form. Previous studies therefore analyzed the model under extremal conditions, yielding solutions that describe a decelerating early-universe phase and asymptotically approach the $\Lambda$CDM model for large values of the scale factor $a(t)$. However, owing to the nonlinearity of Eq.~\eqref{eq:8}, this approach does not permit an explicit determination of the time evolution of the scale factor or the flip time associated with the transition from decelerated to accelerated expansion.

To overcome the limitations, we adopt an approximation methodology by expanding the right-hand side of Eq.~\eqref{eq:8} using a binomial series and retaining only the leading-order terms. This approximation is justified
as the scale factor increases accommodating an expanding universe. Consequently, the truncated equation, presented in Eq.~\eqref{eq:22}, admits an exact analytical solution for the scale factor, which is given in Eq.~\eqref{eq:23}.

On the other hand, for $c<0$, the model admits a phantom-like effective field description and possesses a finite minimum scale factor, leading to a regular cosmological evolution that connects a nonsingular early state to a late-time accelerated epoch. The first-order approximation makes the time evolution explicit and confirms the emergent-universe nature of the solution. The main features are summarized below.

The generalization of the generalized Chaplygin gas model (GCG) which is MCG introduces an additional parameter $A$, the estimated constrained lies within the range $0 < \alpha < \frac{1}{3}$  and $0<A< \frac{1}{3}$ in order to ensure that the speed of sound remains subluminal, preserving causality. Accordingly, our analysis in the paper is carried out within these admissible parameter bounds. Furthermore, cosmological observations indicate that both $\alpha$ and $A$ should take small values in the late universe to achieve an optimal fit with empirical data.
Instead of cosmic time we have rewritten the field equation with redshift parameter $z$. To ensure an accelerating universe at the present epoch, the model requires a positive flip redshift ($z_f>0$), which imposes the constraints $B > \frac{c}{2}(1+3A)(1+A)$ and $\Omega_m < \frac{2}{3(1+A)}$; for the physically relevant range $0 \leq A \leq \frac{1}{3}$, the latter constraint is consistent with current observational estimates of the matter density parameter.

We draw the best-fit curve of the redshift $z$ versus the Hubble parameter $H(z)$ in Fig.~\ref{hz1}, which is obtained using the Hubble $57$ data set. Additionally, Fig.~\ref{hz2} compares this best-fit curve with the theoretical prediction derived from Eq.~\eqref{eq:19}. The excellent agreement between these two curves across the entire range of cosmic evolution demonstrates the consistency of our model with the observational data. Furthermore, we analyze the behavior of the deceleration parameter and its flip time, the effective equation of state and other key cosmological quantities as functions of redshift $z$.

The CMB analysis yields $R_{\mathrm{th}} = 1.734$ with $\chi^2_{\mathrm{CMB}} = 0.9066$, consistent with observations at $z_* \sim 1090$. The combined Hubble-$57$ and CMB analysis provides best-fit values $B_s = 0.695$ and $H_0 = 69.41 \, \mathrm{km\,s^{-1}\,Mpc^{-1}}$, with reduced parameter degeneracy, ensuring consistency across both early and late epochs. Moreover, the small value $A \sim 0.002$ leads to an extended quasi-matter-dominated phase ($w_{\mathrm{eff}} \approx 0$, $\rho \propto a^{-3}$), which supports the standard growth of perturbations $D(a) \propto a$, thereby ensuring compatibility with large-scale structure formation.

In Sec-4, we introduce an alternative formulation where we get the exact solution. For this reason, we can explain the evolution of the scale factor $a(t)$. Also calculate the transition (flip) time from decelerated to accelerated expansion. To get an acceleration at present epoch $z_f> 0$ which yields $\alpha < \frac{1-3A}{3(1+A)}$. We have studied the deceleration parameter and effective equation of state where we note  that the universe initially determined by the parameter $A$ which however finally permits $\Lambda$CDM. Here the CMB analysis gives $R_{\mathrm{th}} = 1.734$ with $\chi^2_{\mathrm{CMB}} = 0.909$, showing good agreement with observations at $z_* \sim 1090$. The combined Hubble-$57$ and CMB analysis yields the best-fit values $B_s = 0.687$ and $H_0 = 69.01 \, \mathrm{km\,s^{-1}\,Mpc^{-1}}$, providing a consistent description across both early and late cosmological epochs.

It is interesting to note that for $c<0$, the Modified Chaplygin Gas model admits a phantom-like effective field description and possesses a finite minimum scale factor, leading to a regular cosmological evolution that connects a nonsingular early state to an asymptotic de Sitter phase. However, the existence of a finite minimum scale factor together with a de Sitter future does not by itself determine the nature of the cosmological evolution, as such behavior may correspond to an emergent universe, a bouncing cosmology, or another nonsingular scenario. Therefore, an explicit determination of the scale-factor evolution is required. Owing to the highly nonlinear nature of the exact evolution equation, Eq.~\eqref{eq:44}, we employ a first-order approximation to obtain an analytical expression for the scale factor. The resulting solution, Eq.~\eqref{eq:46}, shows that the universe approaches a finite, nonzero scale factor asymptotically as $t\rightarrow-\infty$ and subsequently evolves into a late-time accelerated phase. The expansion ultimately approaches the $\Lambda$CDM limit, thereby confirming the emergent-universe nature of the solution within the MCG framework.

It is worth noting that Mukherjee \textit{et al.}~\cite{sm} previously obtained an emergent-universe solution by considering $\alpha = -\tfrac{1}{2}$,
whereas Dutta \textit{et~al.}~\cite{sd} reported that an emergent-type solution does not arise from the conventional MCG model.
In contrast, we note that the cosmological solution, described by Eq.~\eqref{eq:46}, permits  a new class of emergent-type behavior
that naturally originates from the MCG equation of state under the first-order approximation for $c < 0$,
corresponding to the phantom regime, provided the physically admissible conditions
$0 < A < \tfrac{1}{3}$ and $0 < \alpha < 1$ are satisfied.

Furthermore, the present-epoch analysis provides explicit constraints on the model parameters. Within this framework, we have adopted the condition $\beta < 1$, derived in Sec.~5.2. Upon substituting $d = 2(1+\alpha)$, this condition implies $B_s > 0$. From Table~\ref{t5}, we obtain $B_s = 0.91$, which satisfies this inequality and thereby confirms the overall consistency and viability of the model.

\begin{table}[h!]
\centering
\caption{Best-fit parameter values for different formulations of the Modified Chaplygin Gas (MCG) model using Hubble-$57$ and combined Hubble-$57$ + CMB data}
\resizebox{\textwidth}{!}{%
\begin{tabular}{l c c c c c c c c}
\hline
\textbf{Model/Approach} & \textbf{Dataset} & $\chi^2_m$ & $H_0~(\mathrm{km\,s^{-1}\,Mpc^{-1}})$ & $A$ & $\Omega_m$ & $B_s$ & $\alpha$ & Age (Gyr)\\
\hline
MCG  & H$57$ & 46.48 & 72.0 & 0.034 & 0.216 & - & 0.0026 & 13.79 \\
     & H$57$ & 44.35 & 72.0 & 0.025 & - & 0.790 & 0.0280 & 14.054 \\
     & CMB+H$57$ & 21.66 & 69.41 & 0.002 & 0.30 & 0.695 & 0.002 & - \\

Alternative formulation & H$57$ & 44.87 & 71.0 & 0.002 & 0.248 & - & 0.0024 & 13.94 \\
                        & H$57$ & 44.87 & 71.0 & 0.002 & - & 0.7527 & 0.0024 & 13.94 \\
   & CMB+H$57$ & 21.93 & 69.01 & 0.0012 & 0.30 & 0.687 & 0.0024 & - \\

Emergent Type Solution & H$57$ & 84.49 & 80.20 & 0.003 & - & 0.910 & 0.0044 & 12.92 \\
                       & CMB+H$57$ & 65.69 & 83.25 & 0.003 & 0.30 & 0.952 & 0.00135 & - \\
\hline
\end{tabular}%
}
\label{Tdata}
\end{table}

The combined Hubble-$57$ and CMB analysis yields a best-fit value of $B_s = 0.952$, which is close to unity and indicates a late-time dark energy dominated universe. The estimated Hubble parameter, $H_0$, is relatively high but remains consistent with local measurements. This higher value of $H_0$ may also be interpreted in the context of the Hubble tension, highlighting the known discrepancy between early-universe CMB estimates and late-time local measurements.
We further compute the CMB shift parameter, obtaining $R_{\mathrm{th}} = 1.741$ with $\chi^2_{\mathrm{CMB}} = 1.05$ for $z_* \simeq 1090$, which is in good agreement with observations. The inclusion of the CMB constraint significantly tightens the parameter space by incorporating early-universe information, leading to a consistent description of cosmic evolution across both early and late epochs. We also examine the emergent universe solution and find it to be geodesically complete. Furthermore, the finiteness of the Kretschmann scalar confirms the absence of curvature singularities, indicating that the model remains regular throughout the entire cosmic evolution.

It is worth noting that although the Modified Chaplygin Gas (MCG) model has been extensively studied over the past two decades, most investigations have mainly focused on its asymptotic or limiting behaviors. In contrast, the present work provides an exact analytical form of the scale factor within the proposed approximation, allowing a detailed analysis of the cosmic expansion history, including an explicit determination of the flip time. The observational results obtained from both the standard MCG model and the alternative approach, using the Hubble-$57$ and combined Hubble-$57$ \& CMB analyses, are found to be mutually consistent, thereby supporting the model’s ability to describe cosmic evolution from the early universe to the present accelerated epoch.

Although the first-order approximation is formally valid only at late times, its agreement with the combined Hubble-$57$ and CMB analysis over a broad redshift range supports the overall viability of the model.

We further reinterpret the entire analysis within the framework of the Raychaudhuri equation and obtain results that remain consistent with the previous formulations, thereby reinforcing the validity of the adopted approximation and methodology. Table~\ref{Tdata}  summarizes the best-fit parameter values obtained from the Hubble-$57$ and combined Hubble-$57$ \& CMB analyses for all the considered models.

Overall, the consistency among all the approaches demonstrates the robustness of the theoretical framework and indicates that the MCG scenario provides a theoretically sound and observationally consistent explanation for the present cosmic acceleration without invoking additional exotic fields or modifications to general relativity.

\vspace{0.1 cm}
\textbf{Acknowledgement:} We sincerely thank the Referee for the constructive and insightful comments, which have helped us improve the manuscript significantly.\\

\textbf{Declaration of competing interest}
The authors declare that they have no conflict of interest.

\end{document}